\documentclass[11pt]{article}
\usepackage[utf8]{inputenc}
\usepackage[T1]{fontenc}
\usepackage{lmodern}
\usepackage{amsmath,amsthm,amssymb,mathtools}
\usepackage[preprint]{jmlr2e}
\usepackage{booktabs,tabularx,array,microtype,xcolor}
\usepackage{enumitem,placeins,needspace}
\hypersetup{pdftitle={Optimal Polynomial-Query Approximation for Consistent Submodular Maximization with Short Lookahead},pdfauthor={Vaneet Aggarwal},colorlinks=true,linkcolor=blue!40!black,citecolor=blue!40!black,urlcolor=blue!40!black}
\numberwithin{theorem}{section}
\newcommand{\OPT}{\operatorname{OPT}}
\newcommand{\E}{\mathbb E}

\newcommand{\eps}{\varepsilon}

\newenvironment{proofof}[1]{\par\noindent{\bf Proof of #1\ }}{\hfill\BlackBox\\[2mm]}
\setlist{topsep=4pt,itemsep=4pt,parsep=0pt,leftmargin=*}
\allowdisplaybreaks[2]
\title{Optimal Polynomial-Query Approximation for Consistent Submodular Maximization with Short Lookahead}
\author{\name Vaneet Aggarwal\\ \addr Purdue University}
\ShortHeadings{}{}
\firstpageno{1}
\begin{document}
\maketitle
\begin{abstract}
Elements arrive one at a time, and after each arrival an algorithm holds a
set of at most $k$ arrived elements. Its goal is a held set whose value
under a monotone submodular function, for which adding an element never
decreases the value and adds less to a larger set, is close to the best
value that $k$ arrived elements can achieve. The recourse of an update is the number of
elements inserted into or removed from the held set, and we require it to
be bounded independently of $k$ and of the number $n$ of arrivals. If the
algorithm evaluates the function only on sets of arrived elements and
makes polynomially many evaluations, the best ratio a randomized algorithm
can achieve is $2-\sqrt2\approx0.5858$. We show that seeing future
arrivals raises the best ratio for polynomially many evaluations to
$1-1/e\approx0.6321$, for deterministic and for randomized algorithms.
No algorithm with polynomially many evaluations exceeds $1-1/e$, even
when all elements are available at once. For every $\eps>0$, a
deterministic algorithm that sees the next $\lceil\eps k/3\rceil$
arrivals keeps a set of value at least $1-1/e-\eps$ times the optimum at
every time. It changes $O(\eps^{-3})$ elements per arrival
and makes $(n+1)^{O(\eps^{-2})}$ evaluations. With the whole arrival sequence known
in advance, a deterministic algorithm attains the same ratio with
$O(\eps^{-2})$ changes per arrival. A randomized algorithm with the same
short lookahead makes $O(n^2k)$ evaluations, changes
$O(\eps^{-2})$ elements per arrival, and attains the same ratio in
expectation at each fixed time. The main tool is a greedy algorithm with an
explicit guarantee for each of its partial solutions combined with each
initial segment of the known future arrivals. A second tool moves between two such
solutions one element at a time while keeping the value high. The
deterministic algorithm decides when to move by using upper bounds on the
unknown optimum, which it obtains from failed searches for such moves.
\end{abstract}
\section{Introduction}
\label{sec:introduction}

Let $f$ be a nonnegative set function on a finite ground set $V$ with
$f(\varnothing)=0$, where $\varnothing$ is the empty set. It is monotone if $f(A)\le f(B)$ whenever
$A\subseteq B$, and submodular if
$f(A\cup\{e\})-f(A)\ge f(B\cup\{e\})-f(B)$ whenever $A\subseteq B$ and
$e\notin B$, so the gain from adding an element can only shrink as the
set grows.
Coverage functions, where each element covers a weighted collection of
features and a set is worth the total weight it covers, are a standard
example. In consistent submodular maximization \citep{duetting2024}, the
elements of $V$ arrive one at a time in an order $e_1,\ldots,e_n$ that is
fixed in advance. After the $t$-th arrival, an algorithm outputs a set
$S_t$ of at most $k$ elements of $X_t=\{e_1,\ldots,e_t\}$, and $\OPT_t$
denotes the largest value of $f$ on such a set. The recourse at time $t$
is $|S_t\triangle S_{t-1}|$, the number of elements inserted or removed;
it is called constant if it is bounded by a number that may depend on the
target accuracy but not on $k$ or $n$. The algorithm accesses $f$ only
through value queries: it submits a set $A$ and receives the number
$f(A)$. Each query is counted as one unit of time, so the cost of an
algorithm is its number of queries; an algorithm with unrestricted
queries may evaluate $f$ on all $2^n$ sets. A
deterministic algorithm has ratio $c$ if $f(S_t)\ge c\,\OPT_t$ at every
time $t$. A randomized algorithm has ratio $c$ at every fixed time if
$\E f(S_t)\ge c\,\OPT_t$ for every $t$, where the expectation is over its
random choices for a fixed function and arrival order.

The greedy algorithm builds a set by repeatedly adding an element that
increases $f$ the most. Recomputing it after every arrival gives ratio
$q=1-1/e\approx0.6321$ \citep{nemhauser1978}, but it can replace the whole
set at one arrival. With constant recourse and queries on arrived elements
only, \citet{duetting2024} gave a deterministic algorithm with polynomially
many queries and ratio $1/(\varphi+1)\approx0.3820$, where
$\varphi=(1+\sqrt5)/2$, and proved that no deterministic algorithm exceeds
$1/2$, even with unrestricted queries. \Citet{duetting2025}
proved that $2/3$ is the best randomized ratio when the number of queries
is unrestricted, and gave an algorithm with polynomially many queries
and ratio $0.51$.
\Citet{fu2026} proved that $2-\sqrt2\approx0.5858$ is the best randomized
ratio for algorithms that make polynomially many queries on arrived
elements. \Citet{balkanski2026} then gave the algorithm foresight: it
knows the arrival order in advance and may query sets that contain
elements that have not yet arrived, although $S_t$ must still consist of
arrived elements. With unrestricted queries, \citet{balkanski2026}
proved that the
best ratios with foresight are $3/4$ for deterministic algorithms and
approximately $0.7942$ for randomized ones, and asked whether
algorithms with polynomially many queries and constant recourse can
approach $q$. Table~\ref{tab:comparison}
collects these ratios and upper bounds, together with ours.

We answer this question for deterministic and for randomized algorithms.
For every $\eps>0$, a deterministic algorithm that sees only the next
$\lceil\eps k/3\rceil$ arrivals, that is, with lookahead
$\lceil\eps k/3\rceil$, satisfies
\[
 f(S_t)\ge(q-\eps)\OPT_t\qquad\text{at every time }t,
\]
with $O(\eps^{-3})$ recourse and
$(n+1)^{O(\eps^{-2})}$ queries
(Theorem~\ref{thm:deterministic}). With the whole arrival sequence known
in advance, the recourse improves to $O(\eps^{-2})$
(Corollary~\ref{cor:full-foresight}). A randomized algorithm with the same
short lookahead has $O(\eps^{-2})$ recourse and makes $O(n^2k)$
queries, and at each fixed time it attains $\E f(S_t)\ge(q-\eps)\OPT_t$
and has $f(S_t)\ge(q-\eps)\OPT_t$ with probability at least $1-\eps/6$
(Theorem~\ref{thm:randomized}). Every algorithm queries only sets
of at most $k$ elements. No algorithm that makes polynomially many
queries can exceed $q$, even when all elements are available at once
\citep{nemhauserwolsey1978,vondrak2013}. Since this bound also applies
with foresight and unrestricted recourse
(Proposition~\ref{prop:oracle-upper}), $q$ is the best ratio for polynomial-query algorithms with foresight,
deterministic or randomized. The lookahead must grow linearly in
$k$: by a reduction to the barrier of \citet{fu2026}, constant recourse,
polynomially many queries and a ratio above $2-\sqrt2$ together require
lookahead $\Omega(k)$ (Proposition~\ref{prop:lookahead-lower}).
Table~\ref{tab:resources} compares the three algorithms.

\begin{table}[t]
\centering
\small
\renewcommand{\arraystretch}{1.2}
\begin{tabularx}{\textwidth}{@{}>{\raggedright\arraybackslash}X
  >{\raggedright\arraybackslash}p{0.28\textwidth}
  >{\raggedright\arraybackslash}p{0.26\textwidth}@{}}
\toprule
Setting & Best ratio achieved & Upper bound\\
\midrule
\multicolumn{3}{@{}l}{\emph{No limit on recourse, polynomially many queries}}\\
\quad recompute greedy after every arrival
 & $q\approx0.6321$ \citep{nemhauser1978}
 & $q$ \citep{nemhauserwolsey1978,vondrak2013}\\
\midrule
\multicolumn{3}{@{}l}{\emph{Constant recourse, queries on arrived elements only}}\\
\quad deterministic, polynomially many queries
 & $1/(\varphi+1)\approx0.3820$ \citep{duetting2024}
 & $1/2$, even with unrestricted queries \citep{duetting2024}\\
\quad randomized, unrestricted queries
 & $2/3$ \citep{duetting2025}
 & $2/3$ \citep{duetting2025}\\
\quad randomized, polynomially many queries
 & $2-\sqrt2\approx0.5858$ \citep{fu2026}
 & $2-\sqrt2$ \citep{fu2026}\\
\midrule
\multicolumn{3}{@{}l}{\emph{Constant recourse, queries on future elements (foresight)}}\\
\quad deterministic, unrestricted queries
 & $3/4$ \citep{balkanski2026}
 & $3/4$ \citep{balkanski2026}\\
\quad randomized, unrestricted queries
 & $\approx0.7942$ \citep{balkanski2026}
 & $\approx0.7942$ \citep{balkanski2026}\\
\quad deterministic, polynomially many queries
 & $\mathbf q$, Theorem~\ref{thm:deterministic}
 & $q$ \citep{nemhauserwolsey1978,vondrak2013}\\
\quad randomized, polynomially many queries
 & $\mathbf q$, Theorem~\ref{thm:randomized}
 & $q$ \citep{nemhauserwolsey1978,vondrak2013}\\
\bottomrule
\end{tabularx}
\caption{Best ratios and upper bounds for monotone submodular functions
with $f(\varnothing)=0$ under the constraint $|S_t|\le k$, with
$q=1-1/e$ and $\varphi=(1+\sqrt5)/2$. Each ratio is a limit as the
accuracy loss tends to zero. Deterministic ratios hold at every time, and
randomized ratios hold in expectation at each fixed time. Constant
recourse means a bound on $|S_t\triangle S_{t-1}|$ independent of $k$ and
$n$. In the last two rows, polynomially many queries means polynomially
many for each fixed accuracy; Table~\ref{tab:resources} gives the exact
dependence. Proposition~\ref{prop:oracle-upper} shows that the static
upper bound applies with foresight.}
\label{tab:comparison}
\end{table}

\begin{table}[t]
\centering
\small
\renewcommand{\arraystretch}{1.2}
\begin{tabularx}{\textwidth}{@{}>{\raggedright\arraybackslash}X
  >{\raggedright\arraybackslash}p{0.2\textwidth}ccc@{}}
\toprule
Result & Guarantee & Lookahead & Queries & Recourse\\
\midrule
Theorem~\ref{thm:deterministic}, deterministic
 & $f(S_t)\ge(q-\eps)\OPT_t$ at every $t$
 & $\lceil\eps k/3\rceil$ & $(n+1)^{O(\eps^{-2})}$ & $O(\eps^{-3})$\\
Corollary~\ref{cor:full-foresight}, deterministic
 & $f(S_t)\ge(q-\eps)\OPT_t$ at every $t$
 & whole sequence & $(n+1)^{O(\eps^{-3})}$ & $O(\eps^{-2})$\\
Theorem~\ref{thm:randomized}, randomized
 & $\E f(S_t)\ge(q-\eps)\OPT_t$ at each fixed $t$
 & $\lceil\eps k/3\rceil$ & $O(n^2k)$ & $O(\eps^{-2})$\\
\bottomrule
\end{tabularx}
\caption{The three algorithms of this paper. Recourse bounds hold at every
update and, for the randomized algorithm, for every outcome of its random
choices. Every queried set has at most $k$ elements. Neither kind of
algorithm dominates the other: the deterministic guarantees hold at every
time on every run, and the randomized algorithm has smaller recourse with
short lookahead and the query bound $O(n^2k)$, which does not
depend on $\eps$.}
\label{tab:resources}
\end{table}

\paragraph{Approach.}
Except for small budgets $k$, which they handle directly, all three
algorithms use the checkpoint scheme of \citet{balkanski2026}.
At regularly spaced times, called checkpoints, the algorithm picks a set
$A$ of at most
$\kappa$ arrived elements, where the reduced budget $\kappa\le k$ leaves
room for the elements that arrive during the next stretch of time, and
then holds $A$ together with these elements. For this to work,
$f(A\cup R)$ must be large for every
initial segment $R$ of the upcoming arrivals, compared with the best set
of $\kappa$ arrived elements for the same $R$. We call these initial
segments the future prefixes. Different prefixes may have different
optimal sets, so the
classical greedy analysis does not apply directly.

Section~\ref{sec:checkpoint} states this requirement as the checkpoint
subproblem (Definition~\ref{def:checkpoint-subproblem}). Our first tool,
Theorem~\ref{thm:prefix} in Section~\ref{sec:checkpoint}, is a deterministic
greedy algorithm that solves it, and both our deterministic and our
randomized algorithms use it. It keeps computable
upper bounds on the optimal value for each future prefix, divides the
current value for each prefix by its bound, and adds the element that is
best for the latest prefix whose normalized value is nearly the smallest.
Submodularity makes this element at least as useful for every earlier
prefix, and every later prefix already has a larger normalized value.
After $r$ steps, every prefix has at least the fraction
$1-(\kappa/(\kappa+1))^r$ of its optimal value. The randomized checkpoint
schedule of \citet{balkanski2026} moves from one checkpoint set to the
next during a short stretch of updates, called a window, chosen at
random; using our final set in this schedule gives
Theorem~\ref{thm:randomized}.

A deterministic algorithm cannot hide the move from one checkpoint set to
the next inside a random window, so every set it holds during the
move must be good. Lemma~\ref{lem:transition} builds such a move as a
sequence of single-element changes. It adds small batches of elements, found by an
exhaustive search adapted from \citet{udwani2018}, that compensate for
the elements being removed; the guarantee of Theorem~\ref{thm:prefix} for
partial sets is what makes the remaining part of each endpoint valuable
enough. The move is made during a stretch of time in which upper bounds
on the optimum grow slowly; if there is no such stretch, the optimum has
grown so much that the recent arrivals alone are worth enough. The
optimum is unknown, but when a batch search fails, the failure proves an
upper bound on it. Lemma~\ref{lem:time-bounds} turns these upper bounds
into a rule for choosing when to move, and this gives
Theorem~\ref{thm:deterministic}. Section~\ref{sec:algorithm} states both
online algorithms and their guarantees.

\section{Related work}
\label{sec:related}

\paragraph{Maximization without arrivals.}
For a monotone submodular function with $f(\varnothing)=0$, the greedy
algorithm returns a set of at most $k$ elements whose value is at least
$1-(1-1/k)^k\ge1-1/e$ times the optimum \citep{nemhauser1978}. No
algorithm that makes polynomially many value queries does better, even with
randomization \citep{nemhauserwolsey1978,vondrak2013}.

\paragraph{Consistent maximization with queries on arrived elements.}
\Citet{duetting2024} introduced the model, gave the deterministic
Chasing-Local-Opt algorithm with ratio $1/(\varphi+1)$, and proved that no
deterministic algorithm exceeds $1/2$, even with unrestricted queries. \Citet{duetting2025} proved that the
best randomized ratio with unrestricted queries is $2/3$ and gave an
algorithm with polynomially many queries, ratio $0.51$ and constant
recourse. The
reduction of \citet{duetting2025} separates two tasks: choosing a set that remains good after
further elements are added to it, and scheduling the changes between
consecutive such sets. \Citet{fu2026} proved that $2-\sqrt2$ is the best
randomized ratio for polynomially many queries on arrived elements. The
impossibility result of \citet{fu2026} rests on how little the algorithm can learn about the
coming arrivals before an update at which only a few changes are allowed.
Queries on future elements remove this obstacle. For weighted coverage
functions, \citet[Theorem~7.1]{fu2026} obtained ratio $1-1/e-\eps$ with
$O(\eps^{-1})$ recourse using only queries on arrived elements. Our results
hold for every monotone submodular function.

\paragraph{Foresight.}
\Citet{balkanski2026} gave the arrival order and queries on future
elements to the algorithm. With unrestricted queries, the same paper proves that
the best ratios are $3/4$ for deterministic algorithms, with
$O(\eps^{-3}\log(1/\eps))$ recourse, and approximately $0.7942$ for
randomized ones, with $O(\eps^{-2})$ recourse. Both algorithms compute
optimal sets for the future prefixes, which takes exponentially many
queries in general. The randomized one combines the checkpoint reduction
\citep[Lemma~1 and Appendix~A.3]{balkanski2026}, which builds on
\citet[Theorem~3.2]{duetting2025},
with a distribution over these optimal sets. The deterministic one starts
from a certificate, which is a fractional mixture of these optimal sets,
and rounds it to single-element changes by exhaustive search
\citep[Theorem~5]{balkanski2026}; a certificate at level
$1/\sqrt2$ has a short proof \citep[Theorem~6]{balkanski2026}, and the
level $3/4$ is proved with a computer-checked certificate
\citep[Theorem~7]{balkanski2026}. Our deterministic schedule follows the
principle of \citet[Appendix~B.2.4]{balkanski2026}: move between
checkpoint sets during a stretch of time in which the optimum grows
slowly, and otherwise rely on the recent arrivals. The schedule of
\citet{balkanski2026} tests this condition with exact optimal values;
ours uses the upper bounds of Lemma~\ref{lem:time-bounds}, which cost
polynomially many queries for each fixed accuracy.

\paragraph{Deletions and reconfiguration.}
\Citet{duetting2026dynamic} study the fully dynamic setting, in which
elements are also deleted, and give a randomized algorithm with
polynomially many queries that, under a cardinality constraint, has ratio
$1/2-O(\eps)$ in expectation and $O(\eps^{-2})$ recourse.
\Citet{ohsaka2022} study the reconfiguration of a monotone submodular
objective: two sets of $k$ elements must be connected by single-element
exchanges so that every set along the way has large value, and their
polynomial-time algorithm attains half of the smaller endpoint value
\citep[Theorem~5.1]{ohsaka2022}. Lemma~\ref{lem:transition} also connects
two sets through valuable intermediate sets, but its endpoints are
ordered sets with guarantees for all their prefixes, and it compares the
intermediate sets with a common upper bound on the optimum rather than
with the endpoint values.

\paragraph{Several submodular objectives at once.}
\Citet{udwani2018} gave deterministic algorithms for maximizing the
minimum of several monotone submodular functions, where the benchmark is a
single set that is good for all of them. \Citet[Lemma~7]{udwani2018}
shows that such a common set contains a small subset that captures a
proportional share of every objective, and \citet[Theorem~8]{udwani2018}
applies this property batch by batch. Lemma~\ref{lem:small-subset}
proves a statement of this kind with explicit parameters, by rounding a
basic solution of a covering linear program. Theorem~\ref{thm:prefix} uses a different benchmark: each
future prefix has its own optimal set, and the argument relies on the
nesting of the prefixes instead of a common optimal set.

\section{Model}
\label{sec:model}

The ground set is $V=\{e_1,\ldots,e_n\}$, and $f:2^V\to\mathbb R_{\ge0}$
assigns a nonnegative value to every subset of $V$. For a set $A$ and an
element $e$ write $f(e\mid A)=f(A\cup\{e\})-f(A)$ for the marginal value
of $e$ at $A$. The function $f$ is normalized if $f(\varnothing)=0$,
monotone if $f(A)\le f(B)$ whenever $A\subseteq B$, and submodular if
\begin{equation}
 f(e\mid A)\ge f(e\mid B)
 \qquad\text{whenever }A\subseteq B\subseteq V\text{ and }e\in V\setminus B,
 \label{eq:submodularity}
\end{equation}
that is, the marginal value of an element can only decrease as the set
grows. Throughout, $f$ is normalized, monotone and submodular, and
$q=1-1/e$. The algorithm has no other
access to $f$ than value queries: it submits a set $A$ and receives the
number $f(A)$. We say that an algorithm makes polynomially many queries if this number
is polynomial in $n$ for each fixed accuracy $\eps$, and that it has
unrestricted queries if there is no bound on the number, so that it may
evaluate $f$ on every subset of the ground set. We assume that each
query takes one unit of time and measure the cost of an algorithm by its
number of queries. For every algorithm in this paper, the remaining
computation, which consists of comparisons and arithmetic on query
answers, is polynomial in the number of queries and in $n$.

The elements arrive one at a time in the order $e_1,\ldots,e_n$. Time
is counted in arrivals: time $t$ is the moment just after $e_t$ arrives,
and $X_t=\{e_1,\ldots,e_t\}$ is the set of elements that have arrived by
time $t$. After the arrival at time $t$, the algorithm outputs a set
$S_t\subseteq X_t$ with $|S_t|\le k$; we put $S_0=\varnothing$. An element
can be output only after it has arrived, and an element that leaves the
output may return later.

The lookahead $\Lambda\ge0$ of an algorithm is the number of future arrivals it
knows at time $t$: it knows the arrivals up to time
$\min\{t+\Lambda,n\}$ and may query $f$ on any set of elements of
$X_{\min\{t+\Lambda,n\}}$, but it must still choose $S_t$ inside $X_t$. Lookahead $\Lambda=0$ is the setting of \citet{duetting2024}, in which
only arrived elements can be queried. Full foresight, as in
\citet{balkanski2026}, means $\Lambda=n$: the whole order is known in advance.
Apart from what its lookahead reveals, the algorithm does not know $n$ in
advance.
Our algorithms use $\Lambda=\lceil\eps k/3\rceil$, except one variant that
uses full foresight.

The recourse at time $t$ is
\begin{equation}
 C_t=|S_t\triangle S_{t-1}|,
 \label{eq:online-model}
\end{equation}
the number of elements inserted into or deleted from the output at that
update; replacing one element by another costs $2$. We bound the
worst-case recourse $\max_tC_t$. It is called constant if it is bounded
by a number that may depend on the accuracy $\eps$ but not on $n$ or $k$.
Recomputing a good solution from scratch after every arrival can have
recourse as large as $2k$, which is what constant recourse rules out. If
$k\ge n$, keeping every arrival is optimal and changes one element per
update, so we assume $1\le k<n$.

The benchmark at time $t$ is the best value of $k$ arrived elements,
\[
 \OPT_t=\max_{S\subseteq X_t,\ |S|\le k}f(S),
\]
which changes with $t$, and whose optimal sets at different times may
have little in common. A deterministic algorithm has approximation ratio
$c$ if $f(S_t)\ge c\,\OPT_t$ for every $t$, for every $f$ and every
arrival order. A randomized algorithm has approximation ratio $c$ at every
fixed time if $\E f(S_t)\ge c\,\OPT_t$ for every $t$, where the function
and the arrival order are fixed before the algorithm makes its random
choices and the expectation is over these choices only. This is a
guarantee for each time separately: at a given time, some outcomes of the
random choices may give a smaller value. Our results give ratio $q-\eps$
for every $\eps>0$, in both senses.

Except for small budgets, our algorithms hold at most $\kappa\le k$
elements that are chosen together and use the other $k-\kappa$ places for recent arrivals, with
$\kappa=k-O(\eps k)$. The algorithms are analyzed against the benchmark
with this smaller budget,
\[
 \OPT^\kappa_t=\max_{S\subseteq X_t,\ |S|\le\kappa}f(S),
\]
and the loss from the smaller budget is a factor $1-O(\eps)$ by the
following known fact.

\begin{lemma}[{\citealp[proof of Lemma~1]{balkanski2026}}]
\label{lem:cardinality}
For integers $1\le\kappa\le k$ and every $t$,
$\OPT^\kappa_t\ge(\kappa/k)\OPT_t$.
\end{lemma}

\section{The checkpoint subproblem}
\label{sec:checkpoint}

For budgets that are large compared with $1/\eps$, our algorithms use the
checkpoint scheme, which \citet{duetting2025}
introduced without foresight and \citet[Lemma~1 and
Section~3.2]{balkanski2026} adapted to foresight. Split the arrivals into
consecutive blocks $B_1,B_2,\ldots$ of $b$ elements each, and let
$t_j=jb$ be the time at which $B_j$ ends; we call the times $t_j$
checkpoints, and our algorithms take $b=\Theta(\eps k)$. At each
checkpoint $t_j$ the algorithm chooses an ordered set of
at most $\kappa$ elements of $X_{t_j}$, with prefixes
$A_{j,0}\subseteq A_{j,1}\subseteq\cdots\subseteq A_{j,\kappa}$; we call
$A_{j,\kappa}$ the checkpoint set. During block $B_{j+1}$ the algorithm
changes from $A_{j-1,\kappa}$ to $A_{j,\kappa}$, and it holds
$A_{j,\kappa}$ until the next change, during block $B_{j+2}$. Every
arrival after $t_j$ is kept as well, so whenever $A_{j,\kappa}$ is held at
a time $t\le t_j+2b$, the held set contains
$A_{j,\kappa}\cup(X_t\setminus X_{t_j})$. Its size is at most $\kappa+2b$
plus whatever the change from $A_{j-1,\kappa}$ to $A_{j,\kappa}$
temporarily adds.

The checkpoint set $A_{j,\kappa}$ must be fixed at time $t_j$, while it
must be good at all of the next $2b$ times: the arrivals
$X_t\setminus X_{t_j}$ that it is combined with are known in advance but
grow with $t$. In the notation of \citet{balkanski2026}, let $X$ be a
finite set of arrived elements, let $\kappa\ge1$, and let
$\varnothing=R_0\subseteq R_1\subseteq\cdots\subseteq R_{J-1}$ be sets
disjoint from $X$, called the future prefixes. For $S\subseteq X$ define
\begin{equation}
 h_j(S)=f(S\cup R_j),\qquad \ell_j=f(R_j),\qquad
 P_j=\max_{O\subseteq X,\ |O|\le\kappa}h_j(O).
 \label{eq:prefix-objectives}
\end{equation}
The number $P_j$ is the best value that $\kappa$ elements of $X$ reach
together with the prefix $R_j$, and the sets attaining $P_j$ may differ
with $j$. At a checkpoint the algorithm takes $X=X_{t_j}$ and
$R_v=X_{t_j+v}\setminus X_{t_j}$ for $0\le v\le2b$, so $J=2b+1$.

\Citet[Definition~2.1]{balkanski2026} call a possibly random set
$S\subseteq X$ with $|S|\le\kappa$ $\alpha$-addition-robust if
$\E f(S\cup R_j)\ge\alpha P_j$ for every $j$, and
\citet[Lemma~1]{balkanski2026} turns any routine that returns such sets
into a randomized online algorithm with ratio close to $\alpha$; our
randomized algorithm uses this reduction. With unrestricted queries,
\citet[Section~2]{balkanski2026} compute optimal sets $O_j$ attaining
$P_j$ and choose a distribution over them with
$\alpha=\xi^\star\approx0.7942$; the deterministic algorithm of
\citet[Section~3.2]{balkanski2026} also starts from the sets $O_j$.
Computing $O_j$ takes exponentially many queries in general, so neither
routine runs with polynomially many queries. With polynomially many
queries and foresight, the best known ratios before this work were those
of algorithms that ignore foresight, $1/(\varphi+1)$ for deterministic and
$2-\sqrt2$ for randomized algorithms \citep[Figure~1]{balkanski2026}, and
\citet[Section~1.2]{balkanski2026} asked whether such algorithms can
approach $q$.

We ask for a deterministic set, and we also ask that every initial part
of it be good. So that such a set can always be filled to $\kappa$
positions, our algorithms use a fixed pool of $k$ dummy elements, which
never arrive, are never output and do not change values: the value of a
set is the value of its real elements. Internal computations may include
dummy elements, and a query on a set that contains some is answered by
querying its real elements.

\begin{definition}[The checkpoint subproblem]
\label{def:checkpoint-subproblem}
Add $\kappa$ of the dummy elements to $X$, and let $0\le c_1\le\cdots\le c_\kappa\le1$. An
ordered set $x_1,\ldots,x_\kappa$ of distinct elements of $X$ solves the
checkpoint subproblem with the numbers $c_1,\ldots,c_\kappa$ if its
prefixes $A_r=\{x_1,\ldots,x_r\}$ satisfy
\begin{equation}
 f(A_r\cup R_j)\ge c_rP_j\qquad(1\le r\le\kappa,\ 0\le j<J).
 \label{eq:subproblem-target}
\end{equation}
\end{definition}

For $r=\kappa$, condition~\eqref{eq:subproblem-target} says that
$A_\kappa$ is $c_\kappa$-addition-robust. The conditions for $r<\kappa$
strengthen this requirement, and the deterministic algorithm needs
them. It changes from
$A_{j-1,\kappa}$ to $A_{j,\kappa}$ by deleting the elements of
$A_{j-1,\kappa}$ from the end, a batch at a time, while inserting
compensating elements. While $\kappa-s$ of these elements remain, the
remaining elements form the prefix $A_{j-1,\kappa-s}$, whose value must still be a known
fraction $c_{\kappa-s}$ of the optimum. The randomized algorithm uses only
$A_{j,\kappa}$.

In $P_j$, only the chosen set $O$ is restricted to old elements, while the
future prefix $R_j$ is part of the objective. The next lemma compares
$P_j$ with the online benchmark.

\begin{lemma}[{\citealp[proof of Lemma~1]{balkanski2026}}]
\label{lem:prefix-benchmark}
Let $s\le t$ and $R=X_t\setminus X_s$. Then
$\max_{O\subseteq X_s,\,|O|\le\kappa}f(O\cup R)\ge\OPT^\kappa_t$.
\end{lemma}

By monotonicity, if $A\subseteq X_s$ satisfies
$f(A\cup R)\ge c\max_{O\subseteq X_s,\,|O|\le\kappa}f(O\cup R)$, then
$f(A\cup Q)\ge c\,\OPT^\kappa_t$ for every set $Q\supseteq R$.

Since $\OPT^\kappa_t\ge(\kappa/k)\OPT_t$, a solution of the subproblem
with $c_\kappa\approx q$ and $\kappa\approx k$ gives a held set of value
about $q\,\OPT_t$ whenever $A_{j,\kappa}$ is held. For a single prefix
($J=1$), the subproblem is static maximization, and the greedy algorithm
attains $c_r=1-(1-1/\kappa)^r$ for every $r$. For several
prefixes, greedy for one prefix may choose elements that the arrivals in
a later prefix make worthless, and greedy for the last prefix may ignore
what the earlier ones need. We solve the subproblem deterministically
with polynomially many queries, for all prefixes at once, with the
numbers
\begin{equation}
 a_r=1-\left(\frac{\kappa}{\kappa+1}\right)^r
 \qquad(0\le r\le\kappa),
 \label{eq:prefix-curve}
\end{equation}
which are close to the single-prefix greedy curve and give
$a_\kappa\ge q-1/\kappa$.

\begin{lemma}[Comparison with the exponential curve]
\label{lem:curve}
For $0\le r\le\kappa$, $a_r\ge1-e^{-r/\kappa}-1/(\kappa+1)$. In
particular, $a_\kappa\ge q-1/\kappa$.
\end{lemma}

\begin{proof}
Since $\log(1+1/\kappa)=\int_0^{1/\kappa}\frac{dx}{1+x}\ge\frac1{\kappa+1}$,
we have $(\kappa/(\kappa+1))^r=e^{-r\log(1+1/\kappa)}\le e^{-r/(\kappa+1)}$.
The derivative of $e^{-x}$ has absolute value at most one for $x\ge0$, so
by the mean value theorem
$e^{-r/(\kappa+1)}-e^{-r/\kappa}\le\frac r\kappa-\frac r{\kappa+1}
=\frac r{\kappa(\kappa+1)}\le\frac1{\kappa+1}$. Combining the two
inequalities gives $(\kappa/(\kappa+1))^r\le e^{-r/\kappa}+1/(\kappa+1)$,
which is the claim. For $r=\kappa$ it gives
$a_\kappa\ge1-e^{-1}-1/(\kappa+1)\ge q-1/\kappa$.
\end{proof}

Algorithm~1 attains these numbers. Its input is $X$, $\kappa$ and the
prefixes
$R_0\subseteq\cdots\subseteq R_{J-1}$, with the notation
of~\eqref{eq:prefix-objectives}. The algorithm adds one element at a time.
It keeps computable upper bounds $U_j\ge P_j$, measures the progress of
prefix $j$ by $h_j(A_s)/U_j$, and serves the latest prefix whose progress
is nearly the smallest. The $\kappa$ dummy elements of
Definition~\ref{def:checkpoint-subproblem} ensure that every step has an
element to choose; dummy elements are
removed from every reported solution and need no queries.

The bounds $U_j$ start from the value $K$ of ordinary greedy. Run
ordinary greedy on $f$ over $X$: $T_0=\varnothing$, and $T_{s+1}$
adds to $T_s$ an element of largest marginal value. Put
\begin{equation}
 K=\min_{0\le s<\kappa}
 \{f(T_s)+\kappa[f(T_{s+1})-f(T_s)]\}.
 \label{eq:greedy-upper}
\end{equation}
Lemma~\ref{lem:base-greedy} in Appendix~\ref{app:auxiliary} shows that
$P_0\le K\le2P_0$. At a partial set $A_s\subseteq X$ with $|A_s|<\kappa$,
the largest marginal value for prefix $j$ is
\begin{equation}
 d_j(A_s)=\max_{e\in X\setminus A_s} f(e\mid A_s\cup R_j).
 \label{eq:row-marginal}
\end{equation}
The algorithm keeps numbers $U_0,\ldots,U_{J-1}$, initialized to
$U_j=\ell_j+K$. Before each selection it replaces $U_0,\ldots,U_{J-1}$ by
\begin{equation}
 \widehat U_j=\min\{U_j,\ h_j(A_s)+\kappa d_j(A_s)\},\qquad
 U_i=\min_{j\ge i}\widehat U_j.
 \label{eq:suffix-update}
\end{equation}

\begin{lemma}[Upper bounds for the prefix optima]
\label{lem:prefix-bounds}
Suppose $K>0$. Before the first update, $P_i\le U_i$ for every $i$. After every update~\eqref{eq:suffix-update} at a partial set $A_s$,
\begin{enumerate}[label=(\alph*)]
\item $0<P_i\le U_i$ for every $i$;
\item $U_i\le U_{i+1}$ for $i<J-1$;
\item $U_i\le h_i(A_s)+\kappa d_i(A_s)$ for every $i$;
\item each $U_i$ is at most its value before the update.
\end{enumerate}
If $K=0$, then $f(S\cup R_j)=\ell_j=P_j$ for every $S\subseteq X$ and
every $j$.
\end{lemma}

Appendix~\ref{app:prefix-bounds} proves Lemma~\ref{lem:prefix-bounds}.

At step $s$, after the update, the algorithm divides the current value
for each prefix by its bound:
\begin{equation}
 M_s=\min_i\frac{h_i(A_s)}{U_i},\qquad
 \delta_s=\frac{1-M_s}{\kappa+1}.
 \label{eq:minimum-ratio}
\end{equation}
Let $j_s$ be the largest index $j$ with
\begin{equation}
 \frac{h_j(A_s)}{U_j}\le M_s+\delta_s,
 \label{eq:chosen-index}
\end{equation}
and add an element of $X\setminus A_s$ maximizing
$f(e\mid A_s\cup R_{j_s})$, with ties broken by a fixed order.

\begin{center}
\begin{minipage}{0.97\textwidth}
\hrule\vspace{5pt}
\textbf{Algorithm 1. Greedy selection for all future prefixes}
\begin{enumerate}[leftmargin=1.6em,itemsep=3pt]
\item Compute $K$ by~\eqref{eq:greedy-upper}. If $K=0$, return
$\kappa$ dummy elements. Otherwise set $A_0=\varnothing$ and
$U_j=\ell_j+K$ for all $j$.
\item For $s=0,\ldots,\kappa-1$: compute the values
\eqref{eq:row-marginal}, apply~\eqref{eq:suffix-update}, choose $j_s$ by
\eqref{eq:minimum-ratio}--\eqref{eq:chosen-index}, add an element
$x_{s+1}\in\arg\max_{e\in X\setminus A_s}f(e\mid A_s\cup R_{j_s})$, and
put $A_{s+1}=A_s\cup\{x_{s+1}\}$.
\item Return the ordered sequence $x_1,\ldots,x_\kappa$.
\end{enumerate}
\vspace{2pt}\hrule
\end{minipage}
\end{center}

\begin{theorem}[Simultaneous guarantees for all partial sets]
\label{thm:prefix}
Algorithm~1 is deterministic and returns an ordered set $x_1,\ldots,
x_\kappa$ whose prefixes $A_r=\{x_1,\ldots,x_r\}$ satisfy
\begin{equation}
 f(A_r\cup R_j)\ge a_rP_j
 \qquad(0\le r\le\kappa,\ 0\le j<J).
 \label{eq:all-prefix-guarantee}
\end{equation}
In particular, $f(A_\kappa\cup R_j)\ge(q-1/\kappa)P_j$ for every $j$.
The algorithm makes $O(\kappa(|X|+\kappa)J)$ value queries, each on a set
of at most $\kappa+|R_{J-1}|$ elements.
\end{theorem}

\begin{proof}
If $K=0$, Lemma~\ref{lem:prefix-bounds} shows that every set, in
particular every set of dummy elements, attains every $P_j$. Assume
$K>0$. For each step $s<\kappa$, write $U_i$ for the bounds after the
update at step $s$, and $\rho_i=h_i(A_s)/U_i$. Since $|A_s|\le\kappa$, we
have $h_i(A_s)\le P_i\le U_i$ by Lemma~\ref{lem:prefix-bounds}(a), so
$0\le\rho_i\le1$, $0\le M_s\le1$ and $\delta_s\ge0$. An index attaining
$M_s$ satisfies~\eqref{eq:chosen-index}, so $j_s$ exists. Write $j=j_s$
and $x=x_{s+1}$.

We first bound the progress of the chosen prefix. By the choice of $x$,
$f(x\mid A_s\cup R_j)=d_j(A_s)$. Lemma~\ref{lem:prefix-bounds}(c) gives
$U_j\le h_j(A_s)+\kappa d_j(A_s)$, so
$d_j(A_s)\ge(U_j-h_j(A_s))/\kappa$. Dividing by $U_j$ and
using~\eqref{eq:chosen-index},
\begin{equation}
 \frac{f(x\mid A_s\cup R_j)}{U_j}
 \ge\frac{1-\rho_j}{\kappa}
 \ge\frac{1-M_s-\delta_s}{\kappa}
 =\frac{(1-M_s)\bigl(1-\frac1{\kappa+1}\bigr)}{\kappa}
 =\delta_s.
 \label{eq:chosen-progress}
\end{equation}

Next consider a prefix $i\le j$. Then
$A_s\cup R_i\subseteq A_s\cup R_j$, and $x\notin A_s\cup R_j$ because
$x\in X\setminus A_s$ and $X$ is disjoint from $R_j$.
By~\eqref{eq:submodularity},
$f(x\mid A_s\cup R_i)\ge f(x\mid A_s\cup R_j)$. By
Lemma~\ref{lem:prefix-bounds}(b), $0<U_i\le U_j$. Hence
$f(x\mid A_s\cup R_i)/U_i\ge f(x\mid A_s\cup R_j)/U_j\ge\delta_s$
by~\eqref{eq:chosen-progress}, and
\[
 \frac{h_i(A_{s+1})}{U_i}=\rho_i+\frac{f(x\mid A_s\cup R_i)}{U_i}
 \ge M_s+\delta_s.
\]

Now consider a prefix $i>j$. Since $j$ is the largest index
satisfying~\eqref{eq:chosen-index}, $\rho_i>M_s+\delta_s$. By
monotonicity, $h_i(A_{s+1})\ge h_i(A_s)$, so
$h_i(A_{s+1})/U_i>M_s+\delta_s$.

Combining the cases $i\le j$ and $i>j$ gives
$h_i(A_{s+1})\ge(M_s+\delta_s)U_i$ for every $i$, with the bounds $U_i$
of step $s$. If $s+1<\kappa$, the update at step $s+1$ does not increase
any bound, by Lemma~\ref{lem:prefix-bounds}(d), so the normalized values
at step $s+1$ are at least $M_s+\delta_s$ and
\begin{equation}
 1-M_{s+1}\le1-M_s-\delta_s
 =\frac{\kappa}{\kappa+1}(1-M_s).
 \label{eq:minimum-recurrence}
\end{equation}

Finally, since $R_0=\varnothing$ and $A_0=\varnothing$,
$h_0(A_0)=0$ and $M_0=0$. Applying~\eqref{eq:minimum-recurrence} $r$
times gives $1-M_r\le(\kappa/(\kappa+1))^r$, that is, $M_r\ge a_r$, for
$r<\kappa$. For such $r$ and every $j$,
$h_j(A_r)\ge M_rU_j\ge a_rP_j$, using Lemma~\ref{lem:prefix-bounds}(a)
and $a_r\ge0$. For $r=\kappa$, the inequality $h_j(A_{s+1})\ge(M_s+\delta_s)U_j$ with
$s=\kappa-1$ gives
$h_j(A_\kappa)\ge(M_{\kappa-1}+\delta_{\kappa-1})U_j$, and
$1-M_{\kappa-1}-\delta_{\kappa-1}=\frac\kappa{\kappa+1}(1-M_{\kappa-1})
\le(\kappa/(\kappa+1))^\kappa$, so again $h_j(A_\kappa)\ge a_\kappa P_j$.
This proves~\eqref{eq:all-prefix-guarantee}, and Lemma~\ref{lem:curve}
gives $a_\kappa\ge q-1/\kappa$.

Turning to the number of queries, the ordinary greedy run for $K$ makes
$O(\kappa(|X|+\kappa))$ queries on sets of at most $\kappa$ elements, and
the values $\ell_j$ take $J$ queries. At step $s$, the algorithm queries
$h_j(A_s)$ and $h_j(A_s\cup\{e\})$ for each $j<J$ and each real element
$e\in X\setminus A_s$; a dummy element has marginal value zero and needs
no query. There are $\kappa$ steps, so the total is
$O(\kappa(|X|+\kappa)J)$. Every queried set is $S\cup R_j$ with
$|S|\le\kappa$, so it has at most $\kappa+|R_{J-1}|$ elements. The other
computation is polynomial in the number of queries: storing, for each
$j$, the largest marginal found while scanning the elements gives
$d_j(A_s)$, and a backward pass over $j$ computes all suffix minima
in~\eqref{eq:suffix-update}, $M_s$, $\delta_s$ and $j_s$ with $O(J)$
comparisons and arithmetic operations.
\end{proof}

\section{Main results}
\label{sec:algorithm}

For large budgets, both of our online algorithms call Algorithm~1 at
every checkpoint, with
the old set $X_{t_j}$ and the prefixes formed by the next $2b$ arrivals.
The two algorithms differ in how the held set changes from
$A_{j-1,\kappa}$ to $A_{j,\kappa}$. The deterministic algorithm must keep
a good set at every time, including during the change, so it changes
along a path on which every set is valuable, and it must choose when to
change. The randomized algorithm makes the change during a short window
chosen at random, so that a fixed time rarely lies inside the change.

\subsection{Deterministic algorithm}
\label{sec:deterministic}

\citet{balkanski2026} build a valuable path between two checkpoint sets
from a fractional mixture of the exact optimal sets $O_j$ of the
prefixes, and round it to sets by exhaustive search; the time of the
move is chosen by testing how fast the exact optimum grows. Both
procedures use exact optimal sets and values, which take exponentially
many queries in general. Our deterministic algorithm replaces both
procedures. The path is built from the two checkpoint sets themselves, using
the guarantee of Theorem~\ref{thm:prefix} for the partial sets of the
checkpoint sets, and
the time of the move is chosen with upper bounds on the optimum that the
algorithm obtains when its own path construction fails.

Fix a set $Q$ of elements that stays in the
held set during the move, and write $F(S)=f(Q\cup S)$ for $S$ in a set $X$
of available elements. The move starts at an ordered $\kappa$-set with
prefixes $A_0\subseteq\cdots\subseteq A_\kappa$ and ends at an ordered
$\kappa$-set with prefixes $A'_0\subseteq\cdots\subseteq A'_\kappa$, and
$E\ge0$ is a target value. Deleting the elements of $A_\kappa$ one at a
time from the end leaves the prefixes $A_{\kappa-s}$; if
$F(A_r)\ge a_rE$ for every $r$, then after $s$ deletions the remaining
part is still worth $a_{\kappa-s}E$, which is about
$(1-e^{-(\kappa-s)/\kappa})E$. To make up for the deleted part, the
algorithm inserts a set $G_s$ that behaves like $s$ steps of greedy toward
the target: it closes the fraction $1-e^{-s/\kappa}$ of the distance from
$F(A_{\kappa-s})$ to $(1-\eta)E$, where $0<\eta\le1/8$ is an accuracy
parameter. The two fractions combine to
$1-e^{-(\kappa-s)/\kappa}e^{-s/\kappa}=1-1/e$, so every set on the path is
worth about $qE$. The sets $G_s$ are built in batches of $D$ elements, for
the integers
\begin{equation}
 L=\lceil1/\eta\rceil,\qquad m=2(L+1),\qquad D=\lceil m/\eta\rceil,
 \label{eq:batch-parameters}
\end{equation}
so that $m=O(\eta^{-1})$ and $D=O(\eta^{-2})$. Write $z_+=\max\{z,0\}$. A batch $W$ is
accepted at stage $s$ if, for every
fixed set $Y$ among the prefixes $A_r$ and $A'_r$ with $r$ on the grid
$\{\lfloor i\kappa/L\rfloor:0\le i\le L\}$ and $r+s+D\le\kappa$,
\begin{equation}
 F(Y\cup G_s\cup W)-F(Y\cup G_s)
 \ge(1-\eta)\frac D\kappa\bigl((1-\eta)E-F(Y\cup G_s)\bigr)_+ .
 \label{eq:batch-test}
\end{equation}
Testing only the $m$ grid prefixes keeps the number of tests independent
of $\kappa$, and the condition $r+s+D\le\kappa$ keeps every queried set
within $\kappa$ elements.

\begin{center}
\begin{minipage}{0.97\textwidth}
\hrule\vspace{5pt}
\textbf{Algorithm 2. Checked transition}
\begin{enumerate}[leftmargin=1.6em,itemsep=3pt]
\item Set $G_0=\varnothing$ and $s_*=D\lfloor\kappa/D\rfloor$.
\item For $s=0,D,2D,\ldots,s_*-D$: check every $W\subseteq X\setminus G_s$
with $|W|\le D$ in a fixed order, and let $W$ be the first one
satisfying~\eqref{eq:batch-test}. If there is none, report failure and
stop. Otherwise set $G_{s+D}=G_s\cup W$.
\item First branch: for $s=0,D,\ldots,s_*-D$, starting from
$\Gamma_s=A_{\kappa-s}\cup G_s$, insert the elements of
$G_{s+D}\setminus\Gamma_s$ and then delete those of
$\Gamma_s\setminus\Gamma_{s+D}$; finally delete the elements of
$A_{\kappa-s_*}\setminus G_{s_*}$. This leads from $A_\kappa$ to
$G_{s_*}$.
\item Second branch: the same with $A'$ in place of $A$, from $A'_\kappa$
to $G_{s_*}$.
\item Return the first branch followed by the second branch in reverse.
\end{enumerate}
\vspace{2pt}\hrule
\end{minipage}
\end{center}

If some set $O$ of $\kappa$ elements has $F(O)\ge(1-\eta)E$, then a small
subset of $O$ passes all the tests~\eqref{eq:batch-test}
(Lemma~\ref{lem:batch-progress}(a) in Appendix~\ref{app:transition}). A failure
therefore proves an upper bound on the optimum, which Algorithm~3 uses to
time the move.

\begin{lemma}[Checked transition]
\label{lem:transition}
Suppose $\kappa\ge D/\eta$ and $E\ge0$ satisfies
\begin{equation}
 F(A_r)\ge a_rE\quad\text{and}\quad F(A'_r)\ge a_rE
 \qquad(0\le r\le\kappa),
 \label{eq:endpoint-certificate}
\end{equation}
with $a_r$ from~\eqref{eq:prefix-curve}. If Algorithm~2 succeeds, it
returns at most $4\kappa$ single-element insertions and deletions from
$A_\kappa$ to $A'_\kappa$, and every set $S$ it visits satisfies
\begin{equation}
 F(S)\ge(q-5\eta)E\qquad\text{and}\qquad|S|\le\kappa+D.
 \label{eq:transition-value}
\end{equation}
If it fails, then
\begin{equation}
 \max_{O\subseteq X,\ |O|\le\kappa}F(O)<(1-\eta)E.
 \label{eq:failure-certificate}
\end{equation}
It makes $O(m\kappa(|X|+1)^D)$ value queries of $F$, each on a set of at
most $\kappa$ elements of $X$.
\end{lemma}

Appendix~\ref{app:transition} proves Lemma~\ref{lem:transition}.

Algorithm~3, the deterministic online algorithm, follows the checkpoint
scheme of Section~\ref{sec:checkpoint}: at each
checkpoint $t_j$ it computes the next checkpoint set with Algorithm~1, and
during the following block $B_{j+1}$ it moves from $A_{j-1,\kappa}$ to
$A_{j,\kappa}$ along a path returned by Algorithm~2. The remaining
decision is when, within the block, to make the move. Algorithm~3 makes
this decision at time $t_j$, using only the next $2b$ arrivals, by
preparing a move for each of $N$ candidate start times and computing an
upper bound on the optimum at each of them.

Algorithm~3 sets $\eps_0=\min\{\eps,1/2\}$ and $\eta=\eps_0/20$, takes $D$
from~\eqref{eq:batch-parameters}, and lets $K_\eta=\lceil4D/\eta\rceil$. If
$k<K_\eta$, it outputs the greedy set on $X_t$ at every time $t$. For $k\ge K_\eta$ it
uses
\begin{equation}
 b=\lfloor\eta k\rfloor,\qquad
 \kappa=\lfloor(1-3\eta)k\rfloor,\qquad
 N=\lceil8/\eta\rceil+1,\qquad
 w=\lfloor b/N\rfloor.
 \label{eq:det-blocks}
\end{equation}
Thus a block has about $\eta k$ arrivals, a checkpoint set has about
$(1-3\eta)k$ elements, and a block is cut into $N=O(1/\eta)$ intervals of
at least $w$ updates each. During block $B_{j+1}$ the held set is the
union of a core set and a buffer, which until the move ends is
$X_t\setminus X_{t_{j-1}}$ and contains the previous block and the
arrived part of the current block. The core starts at
$A_{j-1,\kappa}$ and moves to $A_{j,\kappa}$ during one interval
$(\tau_i,\tau_{i+1}]$ of the block, where
$\tau_i=t_j+\lfloor ib/N\rfloor$ for $0\le i\le N$. For a candidate start
$\tau=\tau_i$ with $i<N$, the buffer contains the fixed set
\begin{equation}
 Q_\tau=X_\tau\setminus X_{t_{j-1}},\qquad\text{and we put}\quad
 F_\tau(S)=f(Q_\tau\cup S).
 \label{eq:protected-objective}
\end{equation}
Let $K_\tau$ be the number~\eqref{eq:greedy-upper} computed by ordinary
greedy on $X_\tau$ with budget $\kappa$. The search for a move from
$\tau$ starts at
\begin{equation}
 E=\min\left\{K_\tau,\
       \min_{1\le r\le\kappa}\frac{F_\tau(A_{j-1,r})}{a_r},\
       \min_{1\le r\le\kappa}\frac{F_\tau(A_{j,r})}{a_r}\right\},
 \label{eq:initial-transition-bound}
\end{equation}
which satisfies~\eqref{eq:endpoint-certificate} by construction. The
search runs Algorithm~2 and lowers $E$ by the factor $1-\eta$ after each
failure. The initial value is at least $\OPT^\kappa_\tau$, and a failure
at $E$ shows, by~\eqref{eq:failure-certificate}, that
$\OPT^\kappa_\tau<(1-\eta)E$, so the value $E_\tau$ at the first success
is still an upper bound on $\OPT^\kappa_\tau$. We also put $E_{\tau_N}=K_{\tau_N}$ and take the suffix
minima
\begin{equation}
 V_i=\min_{i\le v\le N}E_{\tau_v}\qquad(0\le i\le N),
 \label{eq:time-suffix}
\end{equation}
which are upper bounds that increase with $i$; taking suffix minima uses
the fact that the optimum can only grow with time. The move is made in
the first interval in which these bounds grow slowly, that is, at the
first $i<N$ with
\begin{equation}
 V_{i+1}\le(1+\eta)V_i,
 \label{eq:quiet-interval}
\end{equation}
and in the last interval if there is none. During the chosen interval,
the algorithm carries out the stored sequence of Algorithm~2 for
$\tau_i$, followed by the removal of every element of $B_j$ from
the buffer, spread evenly over the updates of the interval, while
it keeps inserting every new arrival into the buffer.

\begin{center}
\begin{minipage}{0.97\textwidth}
\hrule\vspace{5pt}
\textbf{Algorithm 3. Deterministic algorithm}
\begin{enumerate}[leftmargin=1.6em,itemsep=3pt]
\item Set $\eps_0$, $\eta$, $D$ and $K_\eta$ as above. If $k<K_\eta$, at
every time $t$ output the greedy set of Lemma~\ref{lem:base-greedy} on
$X_t$ with budget $k$, and stop.
\item Set $b$, $\kappa$, $N$ and $w$ by~\eqref{eq:det-blocks}. Let
$A_{0,0}\subseteq\cdots\subseteq A_{0,\kappa}$ consist of dummy elements.
During $B_1$, hold every arrival.
\item At each checkpoint $t_j$, $j\ge1$:
\begin{enumerate}[label=(\alph*),leftmargin=1.4em]
\item run Algorithm~1 on $X_{t_j}$ with budget $\kappa$ and the prefixes
$X_{t_j+v}\setminus X_{t_j}$, $0\le v\le2b$, and call the prefixes of its
output $A_{j,0}\subseteq\cdots\subseteq A_{j,\kappa}$;
\item for $i=0,\ldots,N-1$ and $\tau=\tau_i$: start with $E$
from~\eqref{eq:initial-transition-bound}; run Algorithm~2 with $X$ equal to $X_\tau$ together with the
dummy elements,
$F=F_\tau$ and endpoints $A_{j-1,\cdot}$ and $A_{j,\cdot}$; while it
fails, replace $E$ by $(1-\eta)E$ and run it again; store the returned
sequence and $E_\tau=E$;
\item put $E_{\tau_N}=K_{\tau_N}$, compute~\eqref{eq:time-suffix}, and
choose $i$ by~\eqref{eq:quiet-interval}.
\end{enumerate}
\item During $B_{j+1}$, keep the core and the buffer as separate sets
and hold their union; an element leaves the held set only when it is in
neither. Insert every arrival into the buffer. At each update in
$(\tau_i,\tau_{i+1}]$, also carry out the next
$\lceil(4\kappa+b)/w\rceil$ operations of the list formed by the stored
sequence for $\tau_i$, applied to the core, followed by the
removal of every element of $B_j$ from the buffer; an element that is
also in the core stays held.
\item Output the held set with dummy elements removed.
\end{enumerate}
\vspace{2pt}\hrule
\end{minipage}
\end{center}

The interval $(\tau_i,\tau_{i+1}]$ contains at least $w$ updates, and the
operation list has at most $4\kappa+b$ entries, so the list is finished
within the interval and each update makes $O(\eta^{-2})$ changes. All
computations for block $B_{j+1}$ happen at time $t_j$ and use only the
next $2b$ arrivals.

\begin{lemma}[Upper bounds for choosing the move]
\label{lem:time-bounds}
Let $k\ge K_\eta$ and $j\ge1$. For every candidate start $\tau=\tau_i$
with $i<N$, the search in step~3(b) of Algorithm~3 ends after
$O(1/\eta)$ failures, and
\[
 \OPT^\kappa_\tau\le E_\tau\le2\OPT^\kappa_\tau,\qquad
 F_\tau(S)\ge(q-5\eta)E_\tau
\]
for every set $S$ visited by the stored sequence of operations. Moreover
$\OPT^\kappa_{\tau_i}\le V_i\le2\OPT^\kappa_{\tau_i}$ for $0\le i\le N$,
and $V_0\le V_1\le\cdots\le V_N$.
\end{lemma}

Appendix~\ref{app:time-bounds} proves Lemma~\ref{lem:time-bounds}.
Together with Theorem~\ref{thm:prefix} and Lemma~\ref{lem:prefix-benchmark},
it gives the value of the held set at every time. Before and after the
move, the held set contains a checkpoint set together with the arrivals
since its checkpoint, and Theorem~\ref{thm:prefix} with
Lemma~\ref{lem:prefix-benchmark} gives it value at least
$a_\kappa\OPT^\kappa_t$. During a move in an interval
$(\tau_i,\tau_{i+1}]$ satisfying~\eqref{eq:quiet-interval}, the held set
contains $Q_{\tau_i}$ and a set visited by the stored sequence, so
Lemma~\ref{lem:time-bounds} gives it value at least
$(q-5\eta)E_{\tau_i}\ge(q-5\eta)V_i$, while
$\OPT^\kappa_t\le V_{i+1}\le(1+\eta)V_i$; the ratio is at least
$(q-5\eta)/(1+\eta)\ge q-6\eta$. If~\eqref{eq:quiet-interval} fails for
every $i$, the bounds $V_i$ grow by a factor larger than $8$ before
$\tau_{N-1}$, so the optimum has grown by a factor of at least $4$ since
$t_j$, and the arrivals of the current block alone, which are in the
buffer, are then worth at least $\frac34\OPT^\kappa_t$. The loss
$\OPT^\kappa_t\ge(1-4\eta)\OPT_t$ from the smaller budget gives the
following theorem.

\begin{theorem}[Deterministic algorithm with short lookahead]
\label{thm:deterministic}
For every $0<\eps<1$, Algorithm~3 is deterministic and satisfies, for
every normalized monotone submodular $f$ and every arrival order,
\begin{equation}
 f(S_t)\ge(q-\eps)\OPT_t\quad\text{and}\quad C_t=O(\eps^{-3})
 \qquad\text{for every }t.
 \label{eq:deterministic-guarantee}
\end{equation}
It uses a lookahead of $\lceil\eps k/3\rceil$ and
$(n+1)^{O(\eps^{-2})}$ value queries, each on a set of at most $k$ elements, and its output
always satisfies $S_t\subseteq X_t$ and $|S_t|\le k$.
\end{theorem}

Appendix~\ref{app:det-schedule} proves Theorem~\ref{thm:deterministic}.
The exponent comes from the exhaustive search over batches of
$D=O(\eps^{-2})$ elements in Algorithm~2.

For $k\ge K_\eta$, Algorithm~3 has recourse $O(\eta^{-2})$; the recourse
$O(\eps^{-3})$ in Theorem~\ref{thm:deterministic} comes only
from recomputing greedy when $k<K_\eta=O(\eps^{-3})$. With
full foresight, the algorithm can instead query every set of at most $k$
elements in this case and choose, by dynamic programming, a sequence of
sets of value at least $\frac23\OPT_t$ whose largest change is smallest.
By \citet[Theorems~5 and~6]{balkanski2026}, this largest change is at
most an absolute constant.

\begin{corollary}[Deterministic algorithm with full foresight]
\label{cor:full-foresight}
For every $0<\eps<1$, there is a deterministic full-foresight algorithm
with $f(S_t)\ge(q-\eps)\OPT_t$ and $C_t=O(\eps^{-2})$ at every time. It
uses $(n+1)^{O(\eps^{-3})}$ queries, each on a set of at
most $k$
elements, and its output always satisfies $S_t\subseteq X_t$ and
$|S_t|\le k$.
\end{corollary}

Appendix~\ref{app:full-foresight} proves
Corollary~\ref{cor:full-foresight}.

\subsection{Randomized algorithm}
\label{sec:random-schedule}

The randomized algorithm needs neither a valuable path nor upper bounds
on the optimum. It is the checkpoint reduction of
\citet[Lemma~1 and Appendix~A.3]{balkanski2026}, which builds on
\citet[Theorem~3.2]{duetting2025}, with Algorithm~1 in place of the
routine that computes optimal sets. The new part is therefore only the
routine: Algorithm~1 uses polynomially many queries, while the routine
of \citet{balkanski2026} uses exponentially many.

\begin{center}
\begin{minipage}{0.97\textwidth}
\hrule\vspace{5pt}
\textbf{Algorithm 4. Randomized algorithm}
\begin{enumerate}[leftmargin=1.6em,itemsep=3pt]
\item Set $\eps_0=\min\{\eps,1/2\}$ and $\theta=\eps_0/6$. If
$k<8\theta^{-2}$, at every time $t$ output the greedy set of
Lemma~\ref{lem:base-greedy} on $X_t$ with budget $k$, and stop.
\item Set $b=\lfloor\theta k\rfloor$, $\kappa=k-2b$ and
$n_w=\lceil1/\theta\rceil$. Let $A_{0,\kappa}=\varnothing$. During $B_1$,
hold every arrival.
\item At each checkpoint $t_j$, $j\ge1$: run Algorithm~1 on $X_{t_j}$ with
budget $\kappa$ and the prefixes $X_{t_j+v}\setminus X_{t_j}$,
$0\le v\le2b$, and call its output $A_{j,\kappa}$. Split $B_{j+1}$ into
$n_w$ consecutive windows whose lengths differ by at most one, and choose
one uniformly at random, independently of all other choices.
\item During $B_{j+1}$, hold $T\cup(X_t\setminus X_{t_j})$, where the
temporary set $T$ starts at $A_{j-1,\kappa}\cup B_j$ and becomes
$A_{j,\kappa}$ during the chosen window: pair the elements to be removed
with elements to be added as far as possible, perform the pairs as
exchanges and then the unpaired deletions or insertions, spread as evenly
as possible over the updates of the window.
\end{enumerate}
\vspace{2pt}\hrule
\end{minipage}
\end{center}

Steps~2--4 are Algorithm~1 of \citet[Appendix~A.3]{balkanski2026}, with
Algorithm~1 of this paper as the routine, and step~1 recomputes greedy
instead of an exact optimum for small $k$. At a fixed time $t$, the held
set is a checkpoint set together with the arrivals since its checkpoint
unless $t$ lies in the chosen window, which happens with probability
$1/n_w\le\theta$. Since the window has about $\theta^2k$ updates and the
change involves at most $k$ operations, each update makes
$O(\theta^{-2})$ changes. \Citet{balkanski2026} prove these guarantees
for any routine that returns $\alpha$-addition-robust sets
(Section~\ref{sec:checkpoint}).

\begin{lemma}[{\citealp[Lemma~1 and Appendix~A.3]{balkanski2026}}]
\label{lem:checkpoint}
Fix $0\le\alpha\le1$ and $0<\theta<1/8$, and suppose $k\ge8\theta^{-2}$.
Run steps~2--4 of Algorithm~4 with this $\theta$, where $A_{j,\kappa}$ is
returned by a routine that, given the old set $X_{t_j}$ and the prefixes
of the next $2b$ arrivals, returns an $\alpha$-addition-robust set. Then
$\E f(S_t)\ge(\alpha-3\theta)\OPT_t$ at every fixed time, and every
update changes at most $2\lceil8/\theta^2\rceil+1$ elements.
\end{lemma}

By Theorem~\ref{thm:prefix} and Lemma~\ref{lem:curve}, Algorithm~1 returns
a $(q-1/\kappa)$-addition-robust set, which gives the following
theorem.

\begin{theorem}[Randomized algorithm]
\label{thm:randomized}
For every $0<\eps<1$, Algorithm~4 makes $O(n^2k)$ value queries and
satisfies
\begin{equation}
 \E f(S_t)\ge(q-\eps)\OPT_t\quad\text{for every fixed }t.
 \label{eq:randomized-guarantee}
\end{equation}
For every outcome of its random choices, $C_t=O(\eps^{-2})$ at every
update, $S_t\subseteq X_t$ and $|S_t|\le k$. It uses a lookahead of
$\lceil\eps k/3\rceil$ and queries only sets of at most $k$ elements.
Moreover, at every fixed time $t$,
$\Pr\{f(S_t)\ge(q-\eps)\OPT_t\}\ge1-\eps/6$.
\end{theorem}

Appendix~\ref{app:randomized-theorem} proves Theorem~\ref{thm:randomized}.

\subsection{Optimality of the ratio}
\label{sec:upper}

The ratio $q$ in Theorems~\ref{thm:deterministic} and~\ref{thm:randomized}
cannot be improved with polynomially many queries, even with full
foresight, because an online algorithm with foresight can be simulated by
a static algorithm.

\begin{proposition}[The static upper bound applies with foresight]
\label{prop:oracle-upper}
Fix $\upsilon>0$. No randomized algorithm that makes polynomially many
value queries guarantees $\E f(S_t)\ge(q+\upsilon)\OPT_t$ at every fixed
time on every instance. This holds with full foresight, queries on
arbitrary subsets of $\{e_1,\ldots,e_n\}$ and unrestricted recourse, and
therefore also for deterministic algorithms.
\end{proposition}

\Citet[Figure~1]{balkanski2026} note this bound without proof.
Appendix~\ref{app:oracle-upper} gives the short argument, which runs a
foresight algorithm on the static instance and uses the bound of
\citet{nemhauserwolsey1978} and \citet{vondrak2013}; the guarantee in
expectation at each fixed time needs a step that turns it into a
guarantee with constant probability.

The lookahead in Theorems~\ref{thm:deterministic}
and~\ref{thm:randomized} grows linearly in $k$, and for ratios above
$2-\sqrt2$ the barrier of \citet{fu2026} shows that this order is
necessary.

\begin{proposition}[Lookahead needed for constant recourse]
\label{prop:lookahead-lower}
Fix $\sigma>2-\sqrt2$ and a polynomial $p$. Suppose that for every budget
$k$ some randomized algorithm with lookahead $\Lambda_k$ makes at most
$p(n)$ value queries on every instance with $n$ elements, and satisfies
$|S_t\triangle S_{t-1}|\le C_k$ at every update and
$\E f(S_t)\ge\sigma\OPT_t$ at every fixed time on every instance. Then
there is $c>0$ with $C_k(\Lambda_k+1)\ge ck$ for all sufficiently large
$k$. In particular, if $C_k$ is bounded independently of $k$, then
$\Lambda_k=\Omega(k)$.
\end{proposition}

Appendix~\ref{app:lookahead-lower} proves
Proposition~\ref{prop:lookahead-lower} by inserting zero-value elements
after every arrival, which turns an algorithm with lookahead into one
without lookahead. For
$0<\eps<q-(2-\sqrt2)\approx0.0463$, Theorems~\ref{thm:deterministic}
and~\ref{thm:randomized} have ratio above $2-\sqrt2$ and constant
recourse, so their lookahead has the smallest possible order in $k$.

\section{Conclusion}
\label{sec:conclusion}

With foresight, the best ratio for polynomial-query algorithms with
constant recourse is $1-1/e$, for deterministic and for randomized
algorithms. Our deterministic algorithms use polynomially many queries
for each fixed $\eps$, with $\eps$ in the exponent of $n$.
% Whether a
% deterministic algorithm can attain $1-1/e-\eps$ with constant recourse and
% a number of queries polynomial in $n$, $k$ and $1/\eps$ is open.
Whether the deterministic recourse can be reduced to $O(\eps^{-2})$ with a
lookahead of $\lceil\eps k/3\rceil$ arrivals is open.

\appendix
\section{Standard inequalities and the greedy upper bound}
\label{app:auxiliary}
\label{app:base-greedy}

This appendix first proves two standard inequalities, a bound on a gain
by the sum of marginal values and subadditivity, which are used in
Appendices~\ref{app:prefix}--\ref{app:schedules}. It then proves the
greedy bound and the computable upper bound $K$, which are used in
Sections~\ref{sec:checkpoint} and~\ref{sec:algorithm} and in
Appendices~\ref{app:prefix} and~\ref{app:schedules}.

\begin{lemma}[Sum of marginal values and subadditivity]
\label{lem:basic-inequalities}
Let $F$ be monotone and submodular, and let $A$ and $T$ be sets. Then
\begin{equation}
 F(A\cup T)-F(A)\le\sum_{e\in T\setminus A}F(e\mid A).
 \label{eq:sum-marginals}
\end{equation}
If moreover $F(\varnothing)=0$, then $F$ is subadditive:
\begin{equation}
 F(A\cup T)\le F(A)+F(T).
 \label{eq:subadditive}
\end{equation}
\end{lemma}

\begin{proof}
Add the elements of $T\setminus A=\{x_1,\ldots,x_v\}$ to $A$ one at a
time. The $i$-th increment is $F(x_i\mid A\cup\{x_1,\ldots,x_{i-1}\})$,
which is at most $F(x_i\mid A)$ by~\eqref{eq:submodularity}. Summing the
increments gives~\eqref{eq:sum-marginals}. For~\eqref{eq:subadditive},
write $T=\{y_1,\ldots,y_u\}$ and $T_i=\{y_1,\ldots,y_i\}$. Then
$F(A\cup T)-F(A)=\sum_iF(y_i\mid A\cup T_{i-1})$ and, since
$F(\varnothing)=0$, $F(T)=\sum_iF(y_i\mid T_{i-1})$. Each term of the
first sum is at most the corresponding term of the second
by~\eqref{eq:submodularity}; a term with $y_i\in A$ is zero.
\end{proof}

The greedy recurrence below is the classical argument of
\citet{nemhauser1978}; the upper bound $K$ is read off the same run.

\begin{lemma}[Greedy and a computable upper bound]
\label{lem:base-greedy}
Let $f$ be normalized, monotone and submodular on $X$, and let
$\kappa\ge1$. If $|X|<\kappa$, add zero-value dummy elements to $X$. Run
ordinary greedy for $\kappa$ steps, with prefixes $T_0=\varnothing\subseteq
T_1\subseteq\cdots\subseteq T_\kappa$ and gains
$\gamma_s=f(T_{s+1})-f(T_s)$, and let
\[
 P=\max_{O\subseteq X,\ |O|\le\kappa}f(O),\qquad
 K=\min_{0\le s<\kappa}\{f(T_s)+\kappa\gamma_s\}.
\]
Then $P\le K\le P/q\le2P$ and
\begin{equation}
 f(T_s)\ge[1-(1-1/\kappa)^s]K\qquad(0\le s\le\kappa).
 \label{eq:base-certificate-curve}
\end{equation}
In particular $f(T_\kappa)\ge qP$. If $K=0$, every subset of $X$ has
value zero.
\end{lemma}

\begin{proof}
We first show $P\le K$. Let $O$ attain $P$ and fix $s<\kappa$. By
monotonicity and~\eqref{eq:sum-marginals} with $A=T_s$ and $T=O$,
\[
 P=f(O)\le f(T_s\cup O)\le f(T_s)+\sum_{e\in O\setminus T_s}f(e\mid T_s).
\]
Greedy chose the element of largest marginal value at $T_s$, so each term
of the sum is at most $\gamma_s$, and there are at most $\kappa$ terms.
Hence $P\le f(T_s)+\kappa\gamma_s$, and minimizing over $s$ gives $P\le K$.

For~\eqref{eq:base-certificate-curve}, the definition of $K$ gives
$f(T_s)+\kappa\gamma_s\ge K$, so $\gamma_s\ge(K-f(T_s))/\kappa$ and
\[
 K-f(T_{s+1})=K-f(T_s)-\gamma_s\le(1-1/\kappa)(K-f(T_s)).
\]
Since $f(T_0)=0$, applying this $s$ times gives
$K-f(T_s)\le(1-1/\kappa)^sK$, which is~\eqref{eq:base-certificate-curve}.

For the bound $K\le P/q$, take $s=\kappa$: the inequality
$(1-1/\kappa)^\kappa\le e^{-1}$ and~\eqref{eq:base-certificate-curve} give
$f(T_\kappa)\ge qK$. Since $|T_\kappa|\le\kappa$, $f(T_\kappa)\le P$, so
$qK\le P$. As $q>1/2$, $P/q\le2P$. Together with $P\le K$ this also gives
$f(T_\kappa)\ge qK\ge qP$.

Finally, if $K=0$, then $P=0$, so every singleton has value zero, and
\eqref{eq:sum-marginals} with $A=\varnothing$ shows that every subset of
$X$ has value zero.
\end{proof}

\section{Proof of Lemma~\ref{lem:prefix-bounds} (upper bounds for the prefix optima)}
\label{app:prefix}
\label{app:prefix-bounds}

\begin{proof}
We begin with the bounds before the first update. Let $S\subseteq X$ with $|S|\le\kappa$.
By~\eqref{eq:subadditive}, $f(S\cup R_i)\le f(S)+f(R_i)\le P_0+\ell_i$,
since $h_0(S)=f(S)\le P_0$. Maximizing over $S$ gives $P_i\le P_0+\ell_i$,
and Lemma~\ref{lem:base-greedy} (with $P=P_0$) gives $P_0\le K$, so
$P_i\le\ell_i+K$.

If $K=0$, Lemma~\ref{lem:base-greedy} shows that
every subset of $X$ has value zero. For $S\subseteq X$,
monotonicity and~\eqref{eq:subadditive} give
$\ell_i\le f(S\cup R_i)\le f(S)+\ell_i=\ell_i$. So $f(S\cup R_i)=\ell_i$,
and taking the maximum gives $P_i=\ell_i$. This is the last claim of the
lemma. From now on $K>0$, so $P_0\ge qK>0$ by Lemma~\ref{lem:base-greedy},
and $P_i\ge P_0>0$ by monotonicity (compare $O$ with $O\cup R_i$).

We next show that an update keeps valid upper bounds. Suppose that before an
update at $A_s$ every $U_j$ satisfies $U_j\ge P_j$; by the first paragraph this holds
before the first update. Fix $j$ and let $O_j$ attain $P_j$. By
monotonicity and~\eqref{eq:sum-marginals} with $A=A_s\cup R_j$ and
$T=O_j$,
\[
 P_j=f(O_j\cup R_j)\le f(A_s\cup O_j\cup R_j)
 \le h_j(A_s)+\sum_{e\in O_j\setminus A_s}f(e\mid A_s\cup R_j).
\]
Each element of $O_j\setminus A_s$ lies in $X\setminus A_s$, so each term
is at most $d_j(A_s)$ by~\eqref{eq:row-marginal}, and there are at most
$\kappa$ terms. Hence $P_j\le h_j(A_s)+\kappa d_j(A_s)$, and so
$\widehat U_j\ge P_j$ in~\eqref{eq:suffix-update}.

Properties (a)--(d) follow from this bound and the form
of~\eqref{eq:suffix-update}. (a) The prefixes are nested, so
$P_j\ge P_i$ for $j\ge i$, and the previous paragraph gives $\widehat U_j\ge P_j\ge P_i$.
The minimum over $j\ge i$ is therefore at least $P_i$, which is positive
because $K>0$. (b) The minimum defining $U_{i+1}$ ranges over a subset of the
indices used for $U_i$, so $U_i\le U_{i+1}$. (c) The index $j=i$ is
included, so $U_i\le\widehat U_i\le h_i(A_s)+\kappa d_i(A_s)$. (d) The
same inclusion gives $U_i\le\widehat U_i\le U_i^{\rm old}$, where
$U_i^{\rm old}$ is the value before the update. Since (a) holds after the
update, the previous paragraph applies again at the next update, and induction over the
updates completes the proof.
\end{proof}

\section{Proof of Lemma~\ref{lem:transition} (the checked transition)}
\label{app:transition}

Appendix~\ref{app:small-subset} proves a variant of
\citet[Lemma~7]{udwani2018} with explicit constants, by rounding a
basic solution of a covering linear program. Appendix~\ref{app:residual-progress}
applies it batch by batch, as in \citet[Theorem~8]{udwani2018}, with a
deadline for each fixed set, and Appendix~\ref{app:transition-proof}
builds the path. Throughout this appendix, $0<\eta\le1/8$, and $L$,
$m$ and $D$ are the integers of~\eqref{eq:batch-parameters}.

\subsection{A small batch that improves several objectives at once}
\label{app:small-subset}

\begin{lemma}[A small batch for several objectives]
\label{lem:small-subset}
Let $g_1,\ldots,g_{m'}$, with $m'\le m$, be normalized monotone submodular
functions on a set $X$, and let $\kappa\ge D$. Suppose some $O\subseteq X$
with $|O|\le\kappa$ satisfies $g_i(O)\ge\beta_i\ge0$ for every $i$. Then
some $W\subseteq O$ with $|W|\le D$ satisfies
\begin{equation}
 g_i(W)\ge(1-\eta)\frac D\kappa\beta_i\qquad(1\le i\le m').
 \label{eq:small-subset-bound}
\end{equation}
Checking every subset of $X$ with at most $D$ elements finds such a set
with $O(m(|X|+1)^D)$ value queries.
\end{lemma}

\begin{proof}
If $m'=0$ there is nothing to prove. If $|O|\le D$, take $W=O$; then
$g_i(W)\ge\beta_i\ge(1-\eta)(D/\kappa)\beta_i$ because $D\le\kappa$.
Assume from now on that $v=|O|>D$, and list the elements of $O$ in an
arbitrary fixed order $o_1,\ldots,o_v$, with $O_j=\{o_1,\ldots,o_j\}$ and
$O_0=\varnothing$. For each $i$ and $j$ put
\[
 w_{ij}=g_i(O_j)-g_i(O_{j-1}).
\]
Monotonicity gives $w_{ij}\ge0$, and since $g_i$ is normalized, the sum
telescopes to $\sum_{j=1}^vw_{ij}=g_i(O)$.

We first compare the value of a subset of $O$ with these weights. Let
$W\subseteq O$, and list its elements in the order of $O$ as
$o_{j_1},\ldots,o_{j_u}$ with $j_1<\cdots<j_u$. Since $g_i$ is
normalized, $g_i(W)$ is the sum of the gains
$g_i(o_{j_l}\mid\{o_{j_1},\ldots,o_{j_{l-1}}\})$ for $l=1,\ldots,u$. The
set $\{o_{j_1},\ldots,o_{j_{l-1}}\}$ is contained in $O_{j_l-1}$, so
by~\eqref{eq:submodularity} each gain is at least
$g_i(o_{j_l}\mid O_{j_l-1})=w_{ij_l}$. Hence
\begin{equation}
 g_i(W)\ge\sum_{o_j\in W}w_{ij}\qquad(W\subseteq O).
 \label{eq:chain-bound}
\end{equation}

Next we choose $W$ through a linear program. Put
$\lambda=(D-m'+1)/v$. Since $1\le m'\le m\le D<v$, we have
$0<\lambda<1$. Consider
\begin{equation}
 \begin{aligned}
 &\text{minimize }\sum_{j=1}^vx_j\\
 &\text{subject to }\sum_{j=1}^vw_{ij}x_j\ge\lambda g_i(O)\ \ (1\le i\le m')
 \ \text{ and }\ 0\le x_j\le1\ \ (1\le j\le v).
 \end{aligned}
 \label{eq:shrink-lp}
\end{equation}
The vector with every $x_j=\lambda$ is feasible, because
$\sum_jw_{ij}\lambda=\lambda g_i(O)$, and its objective value is
$\lambda v=D-m'+1$. The feasible region is nonempty and bounded, so the
minimum is attained at a vertex $x^*$, and the minimum is at most
$D-m'+1$. At a vertex of a polyhedron in $\mathbb R^v$, some $v$ linearly
independent constraints hold with equality. A coordinate with
$0<x^*_j<1$ makes neither of its two bound constraints tight, and a
coordinate with $x^*_j\in\{0,1\}$ makes exactly one of them tight. If $F$
coordinates of $x^*$ are fractional, the tight constraints therefore
consist of at most $v-F$ bound constraints and at most $m'$ covering
constraints, so $v\le(v-F)+m'$, that is, $F\le m'$.

Let $W=\{o_j:x^*_j>0\}$. Its indicator vector is at least $x^*$ in every
coordinate, so, as $w_{ij}\ge0$, \eqref{eq:chain-bound} and the covering
constraints give
\[
 g_i(W)\ge\sum_{o_j\in W}w_{ij}\ge\sum_{j=1}^vw_{ij}x^*_j\ge\lambda g_i(O).
\]
For the size, $|W|=\sum_jx^*_j+\sum_j(1-x^*_j)$, where the second sum
runs over the fractional coordinates. If $F=0$, then
$|W|=\sum_jx^*_j\le D-m'+1\le D$. If $F\ge1$, each term of the second
sum is less than $1$, so $|W|<(D-m'+1)+F\le D+1$, and since $|W|$ is an
integer, $|W|\le D$.

Finally, $v\le\kappa$ and $m'\le m$ give
$\lambda\ge(D-m)/\kappa$, and $D\ge m/\eta$ by~\eqref{eq:batch-parameters}
gives $D-m\ge(1-\eta)D$. Hence
$g_i(W)\ge(1-\eta)(D/\kappa)g_i(O)\ge(1-\eta)(D/\kappa)\beta_i$, which
is~\eqref{eq:small-subset-bound}. The linear program is used only in
this proof; the algorithm does not know $O$. For the search, list each
subset of $X$ with at most $D$ elements as a sequence of $D$ symbols
from $X$ and a blank symbol; there are at most $(|X|+1)^D$ subsets, and
each is checked with $O(m)$ queries.
\end{proof}

\subsection{Progress of the batch sequence}
\label{app:residual-progress}

\begin{lemma}[Progress of the batch sequence]
\label{lem:batch-progress}
Let $F$ be nonnegative, monotone and submodular on $X$, not necessarily
normalized, and let $\kappa\ge D$. Fix at most $m$ sets $Y_i\subseteq X$
with $|Y_i|\le r_i\le\kappa$, and a number $T\ge0$. Let $G_0=\varnothing$.
At each stage $s$, a multiple of $D$ with $s+D\le\kappa$, call $i$
\emph{active} if $r_i+s+D\le\kappa$, check every
$W\subseteq X\setminus G_s$ with $|W|\le D$ for
\begin{equation}
 F(Y_i\cup G_s\cup W)-F(Y_i\cup G_s)
 \ge(1-\eta)\frac D\kappa\bigl(T-F(Y_i\cup G_s)\bigr)_+
 \quad\text{for every active }i,
 \label{eq:residual-test}
\end{equation}
and set $G_{s+D}=G_s\cup W$ for the first $W$, in a fixed order, that
passes; if none passes, stop and report failure. Then:
\begin{enumerate}[label=(\alph*)]
\item if some $O\subseteq X$ with $|O|\le\kappa$ has $F(O)\ge T$, no stage
fails;
\item after successful stages up to $s$, for every $i$ with
$r_i+s\le\kappa$,
\begin{equation}
 F(Y_i\cup G_s)\ge(1-z_s)T+z_sF(Y_i),\qquad z_s=e^{-(1-\eta)s/\kappa};
 \label{eq:batch-curve}
\end{equation}
\item every set at which $F$ is queried has at most $\kappa$ elements.
\end{enumerate}
\end{lemma}

\begin{proof}
\emph{(a)} Fix a stage $s$ and, for each active $i$, define
$g_i(W)=F(Y_i\cup G_s\cup W)-F(Y_i\cup G_s)$ for
$W\subseteq X\setminus G_s$. Each $g_i$ is normalized, monotone and
submodular. By monotonicity of $F$,
\[
 g_i(O\setminus G_s)=F(Y_i\cup G_s\cup O)-F(Y_i\cup G_s)
 \ge T-F(Y_i\cup G_s),
\]
and $g_i(O\setminus G_s)\ge0$, so $g_i(O\setminus G_s)\ge
\beta_i:=(T-F(Y_i\cup G_s))_+$. The set $O\setminus G_s$ has at most
$\kappa$ elements, and there are at most $m$ active indices, so
Lemma~\ref{lem:small-subset} on the ground set $X\setminus G_s$ yields a
$W$ with $|W|\le D$ that satisfies~\eqref{eq:residual-test}. Hence the
stage does not fail.

\emph{(b)} Fix $i$ with $r_i+s\le\kappa$ and put
$\Delta_u=(T-F(Y_i\cup G_u))_+$ for multiples $u$ of $D$. For every stage
$u<s$ we have $r_i+u+D\le r_i+s\le\kappa$, so $i$ was active at $u$. If
$\Delta_u=0$, then $F(Y_i\cup G_{u+D})\ge F(Y_i\cup G_u)\ge T$ and
$\Delta_{u+D}=0$. Otherwise~\eqref{eq:residual-test} gives
\[
 T-F(Y_i\cup G_{u+D})\le\Bigl(1-(1-\eta)\frac D\kappa\Bigr)
 \bigl(T-F(Y_i\cup G_u)\bigr).
\]
The factor in parentheses lies in $[0,1]$, so in both cases
$\Delta_{u+D}\le(1-(1-\eta)D/\kappa)\Delta_u$. Applying this $s/D$ times
and using $1-x\le e^{-x}$,
\[
 \Delta_s\le\Bigl(1-(1-\eta)\frac D\kappa\Bigr)^{s/D}\Delta_0
 \le z_s\,(T-F(Y_i))_+.
\]
If $F(Y_i)\le T$, this reads $T-F(Y_i\cup G_s)\le z_s(T-F(Y_i))$, which
rearranges to~\eqref{eq:batch-curve}. If $F(Y_i)>T$, the right-hand side
of~\eqref{eq:batch-curve} is at most $F(Y_i)$, which is at most
$F(Y_i\cup G_s)$ by monotonicity.

\emph{(c)} Each batch has at most $D$ elements, so $|G_s|\le s$. For an
active $i$ and a candidate $W$,
$|Y_i\cup G_s\cup W|\le r_i+s+D\le\kappa$, and the set $Y_i\cup G_s$ is
smaller.
\end{proof}

\subsection{Proof of Lemma~\ref{lem:transition}}
\label{app:transition-proof}

\begin{proof}
Put $T=(1-\eta)E$ and
let $\mathcal R=\{\lfloor i\kappa/L\rfloor:0\le i\le L\}$. Consecutive
elements of $\mathcal R$ differ by at most $\lceil\kappa/L\rceil$, and
$\kappa/L\le\eta\kappa$ because $L\ge1/\eta$. Hence for every integer
$0\le u\le\kappa$, the largest $r\in\mathcal R$ with $r\le u$ satisfies
\begin{equation}
 u-r\le\eta\kappa+1.
 \label{eq:grid-gap}
\end{equation}
The procedure runs the construction of Lemma~\ref{lem:batch-progress} with
target $T$ and the at most $m=2(L+1)$ fixed sets $A_r$ and $A'_r$,
$r\in\mathcal R$, with $r_i=r$. The assumption $\kappa\ge D/\eta$
implies $\kappa\ge D$, the hypothesis of
Lemmas~\ref{lem:small-subset} and~\ref{lem:batch-progress}. If a stage
fails, Lemma~\ref{lem:batch-progress}(a) shows
that no $O\subseteq X$ with $|O|\le\kappa$ has $F(O)\ge T=(1-\eta)E$,
which is~\eqref{eq:failure-certificate}. From now on every stage
succeeds; let $s_*=D\lfloor\kappa/D\rfloor$ be the last stage index, so
$\kappa-D<s_*\le\kappa$.

For each multiple $s$ of $D$ with
$s\le s_*$ put
\begin{equation}
 \Gamma_s=A_{\kappa-s}\cup G_s.
 \label{eq:branch-states}
\end{equation}
Let $r$ be the largest element of $\mathcal R$ with $r\le\kappa-s$. Since
$r+s\le\kappa$, Lemma~\ref{lem:batch-progress}(b) applies to $Y_i=A_r$.
Using monotonicity ($A_r\subseteq A_{\kappa-s}$), \eqref{eq:batch-curve},
\eqref{eq:endpoint-certificate}, and Lemma~\ref{lem:curve},
\begin{align}
 F(\Gamma_s)&\ge F(A_r\cup G_s)
 \ge(1-z_s)(1-\eta)E+z_sF(A_r)\nonumber\\
 &\ge\bigl[(1-z_s)(1-\eta)+z_s(1-e^{-r/\kappa}-1/\kappa)\bigr]E
 \ge\bigl[1-e^{-((1-\eta)s+r)/\kappa}-\eta-1/\kappa\bigr]E,
 \label{eq:full-path-bound}
\end{align}
where the last step expands the bracket, drops the term $z_s\eta\ge0$,
replaces $z_s/\kappa$ by the larger $1/\kappa$, and uses
$z_se^{-r/\kappa}=e^{-((1-\eta)s+r)/\kappa}$.

The exponent in~\eqref{eq:full-path-bound} is controlled
by~\eqref{eq:grid-gap}: with $u=\kappa-s$ and $s\le\kappa$,
\[
 (1-\eta)s+r\ge(1-\eta)s+\kappa-s-\eta\kappa-1
 =\kappa-\eta s-\eta\kappa-1\ge(1-2\eta)\kappa-1.
\]
So the exponent in~\eqref{eq:full-path-bound} is at least $1-x$ with
$x=2\eta+1/\kappa$. The assumption $\kappa\ge D/\eta$ and $D\ge1$ give
$1/\kappa\le\eta$, so $x\le3\eta$. For $0\le x\le1$,
\begin{equation}
 e^{-1+x}-e^{-1}=\int_0^xe^{-1+y}\,dy\le x,
 \label{eq:exp-shift}
\end{equation}
so $e^{-((1-\eta)s+r)/\kappa}\le e^{-1}+2\eta+1/\kappa$.
Substituting into~\eqref{eq:full-path-bound},
\begin{equation}
 F(\Gamma_s)\ge(q-3\eta-2/\kappa)E\ge(q-5\eta)E.
 \label{eq:branch-values}
\end{equation}

The last auxiliary set $G_{s_*}$ is also valuable. The fixed set $A_0=\varnothing$ has
$r=0\in\mathcal R$ and $0+s_*\le\kappa$, so~\eqref{eq:batch-curve} and
$F(\varnothing)\ge0$ give $F(G_{s_*})\ge(1-z_{s_*})(1-\eta)E$. Since
$s_*>\kappa-D$, we have $(1-\eta)s_*/\kappa\ge1-\eta-D/\kappa$, and
by~\eqref{eq:exp-shift} with $x=\eta+D/\kappa\le2\eta$,
\begin{equation}
 F(G_{s_*})\ge(1-\eta)(q-\eta-D/\kappa)E\ge(q-2\eta-D/\kappa)E
 \ge(q-3\eta)E.
 \label{eq:last-aux-value}
\end{equation}

We now build the path. Its first branch goes from
$\Gamma_0=A_\kappa$ through $\Gamma_D,\Gamma_{2D},\ldots,\Gamma_{s_*}$ to
$G_{s_*}$. To go from $\Gamma_s$ to $\Gamma_{s+D}$, first insert the
elements of $G_{s+D}\setminus\Gamma_s$ one at a time, then delete those of
$\Gamma_s\setminus\Gamma_{s+D}$ one at a time. Every set during the
insertions contains $\Gamma_s$, and every set during the deletions
contains $\Gamma_{s+D}$, so by monotonicity
and~\eqref{eq:branch-values} each has value at least $(q-5\eta)E$. Since
$G_s\subseteq G_{s+D}$, the deleted elements lie in
$A_{\kappa-s}\setminus A_{\kappa-s-D}$, so there are at most $D$
deletions, and there are at most $D$ insertions. Finally, delete the
elements of $A_{\kappa-s_*}\setminus G_{s_*}$; every set during these
deletions contains $G_{s_*}$, so~\eqref{eq:last-aux-value} bounds its value. Every set has at
most $|\Gamma_s|+D\le\kappa+D$ elements, because $|\Gamma_s|\le(\kappa-s)+s$.
The branch makes at most $2D$ changes per stage and at most
$\kappa-s_*$ final deletions, in total at most
$2s_*+(\kappa-s_*)\le2\kappa$.

The construction and the bounds above apply
verbatim to $\Gamma'_s=A'_{\kappa-s}\cup G_s$, with the same sets $G_s$,
because the sets $A'_r$, $r\in\mathcal R$, were tested at every stage.
This gives a branch from $A'_\kappa$ to the same set $G_{s_*}$. The
procedure returns the first branch followed by the second in reverse. It
visits only sets already bounded above, uses at most $4\kappa$ changes,
and satisfies~\eqref{eq:transition-value}.

Finally, there are at most $\kappa/D$ stages. Each checks at
most $(|X|+1)^D$ candidates against at most $m$ fixed sets, with $O(1)$
queries per check, so the total is $O(m\kappa(|X|+1)^D)$. By
Lemma~\ref{lem:batch-progress}(c), each query of $F$ is on at most
$\kappa$ elements.
\end{proof}

\section{Proofs of Lemma~\ref{lem:time-bounds}, Theorem~\ref{thm:deterministic}, Corollary~\ref{cor:full-foresight}, Theorem~\ref{thm:randomized} and Propositions~\ref{prop:oracle-upper} and~\ref{prop:lookahead-lower}}
\label{app:schedules}

Throughout this appendix, $\OPT^\kappa_t$ is defined in
Section~\ref{sec:model}. It is nondecreasing in $t$, because $X_t$
grows. The subsections follow the order of Section~\ref{sec:algorithm}; the
proof of Lemma~\ref{lem:checkpoint} is in \citet[Appendix~A.3]{balkanski2026}.

\subsection{Proof of Lemma~\ref{lem:time-bounds}}
\label{app:time-bounds}

This subsection and the next use the parameters of Algorithm~3 with
$k\ge K_\eta$. Since $\eta=\eps_0/20\le1/40$ and $D\ge2/\eta^2$, the assumption
$k\ge K_\eta\ge4D/\eta$ gives the following. First, $D\le\eta k/4$ and
$1/k\le\eta$. Second, $\kappa\ge(1-3\eta)k-1\ge3.7D/\eta-1\ge D/\eta$, so
also $1/\kappa\le\eta$. Third, $\eta k\ge2$, so $b\ge\eta k-1\ge\eta k/2$;
and $N\le8/\eta+2\le9/\eta$, so $b/N\ge\eta^2k/18\ge2$ and
\begin{equation}
 w=\lfloor b/N\rfloor\ge\frac b{2N}\ge\frac{\eta^2k}{36}.
 \label{eq:det-interval-length}
\end{equation}
Fourth,
\begin{equation}
 \kappa+D+2b\le(1-3\eta)k+\eta k/4+2\eta k\le k.
 \label{eq:det-capacity}
\end{equation}

\begin{proof}
Fix $j\ge1$ and a candidate start $\tau=\tau_i$ with $i<N$, so
$t_j\le\tau\le t_{j+1}$.

We first show that for $0\le r\le\kappa$,
\begin{equation}
 F_\tau(A_{j-1,r})\ge a_r\OPT^\kappa_\tau\quad\text{and}\quad
 F_\tau(A_{j,r})\ge a_r\OPT^\kappa_\tau.
 \label{eq:det-both-endpoints}
\end{equation}
Consider first the new prefixes $A_{j,r}$, computed at $t_j$, and the prefix
$R=X_\tau\setminus X_{t_j}$, which has $\tau-t_j\le b$ elements and was
supplied to Algorithm~1. Theorem~\ref{thm:prefix} gives
$f(A_{j,r}\cup R)\ge a_r\max_{O\subseteq X_{t_j},\,|O|\le\kappa}f(O\cup R)$,
and Lemma~\ref{lem:prefix-benchmark} with $s=t_j$ and $t=\tau$, together
with monotonicity and $Q_\tau\supseteq R$, gives
$F_\tau(A_{j,r})\ge a_r\OPT^\kappa_\tau$. For
the old prefixes $A_{j-1,r}$ with $j\ge2$, the same argument applies at
$t_{j-1}$ with the prefix $Q_\tau=X_\tau\setminus X_{t_{j-1}}$, which has
$\tau-t_{j-1}\le2b$ elements and was supplied at $t_{j-1}$. For $j=1$,
the set $A_{0,r}$ consists of dummy elements and $Q_\tau=X_\tau$, so
$F_\tau(A_{0,r})=f(X_\tau)\ge\OPT^\kappa_\tau\ge a_r\OPT^\kappa_\tau$.

Next we bound the initial value of $E$. Lemma~\ref{lem:base-greedy}
applied to $f$ on $X_\tau$ with budget $\kappa$ gives
$\OPT^\kappa_\tau\le K_\tau\le2\OPT^\kappa_\tau$. By~\eqref{eq:det-both-endpoints}, each ratio
in~\eqref{eq:initial-transition-bound} is at least $\OPT^\kappa_\tau$.
Hence the initial $E$ satisfies
\begin{equation}
 \OPT^\kappa_\tau\le E\le K_\tau\le2\OPT^\kappa_\tau.
 \label{eq:det-upper-invariant}
\end{equation}
By the definition of $E$ as a minimum, $F_\tau(A_{j-1,r})\ge a_rE$ and
$F_\tau(A_{j,r})\ge a_rE$ for $r\ge1$, and for $r=0$ these hold because
$a_0=0$ and $F_\tau\ge0$. So~\eqref{eq:endpoint-certificate} holds, and
it continues to hold when $E$ is decreased.

Each failure preserves~\eqref{eq:det-upper-invariant}. The
assumption $\kappa\ge D/\eta$ of Lemma~\ref{lem:transition} holds by the
inequalities at the start of this section. If the procedure fails at the
current $E$, then by~\eqref{eq:failure-certificate} and monotonicity,
\[
 \OPT^\kappa_\tau\le\max_{O\subseteq X_\tau,\,|O|\le\kappa}F_\tau(O)
 <(1-\eta)E,
\]
so the new value $(1-\eta)E$ still satisfies~\eqref{eq:det-upper-invariant}.

The number of failures is therefore bounded. After $u$ failures,
$E\le(1-\eta)^uK_\tau\le2e^{-\eta u}\OPT^\kappa_\tau$. Once
$u\ge\ln2/\eta-1$, the next target satisfies
$(1-\eta)E\le2e^{-\eta(u+1)}\OPT^\kappa_\tau\le\OPT^\kappa_\tau$, and by
the previous paragraph a failure would give $\OPT^\kappa_\tau<(1-\eta)E\le
\OPT^\kappa_\tau$, which is impossible. So there are at most
$\lceil\ln2/\eta\rceil=O(1/\eta)$ failures. (If $\OPT^\kappa_\tau=0$,
then $E=0$, and a failure would give a negative maximum, so the first
attempt succeeds.)

The final value $E_\tau$
satisfies~\eqref{eq:det-upper-invariant}, which gives
$\OPT^\kappa_\tau\le E_\tau\le2\OPT^\kappa_\tau$, and the successful
attempt gives $F_\tau(S)\ge(q-5\eta)E_\tau$ for every visited set $S$
by~\eqref{eq:transition-value}.

For the suffix minima, Lemma~\ref{lem:base-greedy} first gives, at
$\tau_N$, $\OPT^\kappa_{\tau_N}\le E_{\tau_N}=K_{\tau_N}\le
2\OPT^\kappa_{\tau_N}$. So for every $v$, $E_{\tau_v}\ge
\OPT^\kappa_{\tau_v}$, and for $v\ge i$ this is at least
$\OPT^\kappa_{\tau_i}$ because $\OPT^\kappa_t$ is nondecreasing. Taking
the minimum in~\eqref{eq:time-suffix} gives $V_i\ge\OPT^\kappa_{\tau_i}$,
and the term $v=i$ gives $V_i\le E_{\tau_i}\le2\OPT^\kappa_{\tau_i}$.
Finally $V_i=\min\{E_{\tau_i},V_{i+1}\}\le V_{i+1}$.
\end{proof}

\subsection{Proof of Theorem~\ref{thm:deterministic}}
\label{app:det-schedule}

\begin{proof}
If $k<K_\eta$, the algorithm outputs the
greedy set of Lemma~\ref{lem:base-greedy} on $X_t$ with budget $k$, with
dummy elements removed. By that lemma, $f(S_t)\ge q\OPT_t$. Two sets of at
most $k$ elements differ in at most
$2k<2K_\eta=O(\eta^{-3})$ elements. Greedy on $X_t$ uses
polynomially many queries, all on sets of at most $k$ elements, and no
lookahead. Assume $k\ge K_\eta$ for the rest of the proof. At the
end of the sequence we append zero-value padding arrivals to complete the
current block and two further blocks, extend $f$ by
$\widetilde f(S)=f(S\cap\{e_1,\ldots,e_n\})$, which is normalized,
monotone and submodular and needs no new queries, and report the outputs
at real times with padding and dummy elements removed. This changes neither values
nor the bounds on $|S_t|$ and $C_t$.

We first describe the held sets. During $B_1$ the algorithm holds every
arrival; $|S_t|\le b<k$ and $S_t=X_t$ is optimal. During block $B_{j+1}$,
$j\ge1$, the held set is $S_t=\Gamma\cup H_t$, where the buffer
$H_t$ equals $X_t\setminus X_{t_{j-1}}$ until the buffer removals start
and lies between $X_t\setminus X_{t_j}$ and $X_t\setminus X_{t_{j-1}}$
afterwards, and where
the core $\Gamma$ is: $A_{j-1,\kappa}$ up to the start $\tau_i$ of the
chosen interval; the sets visited by the stored sequence for $\tau_i$
during the interval; and $A_{j,\kappa}$ afterwards. After the
core reaches $A_{j,\kappa}$, the buffer removals take every element of
$B_j$ out of the buffer, so that afterwards $H_t=X_t\setminus X_{t_j}$ and
the held set is $A_{j,\kappa}\cup(X_t\setminus X_{t_j})$. At $t_{j+1}$ this is
$A_{j,\kappa}\cup B_{j+1}$, the starting state for the next block, so no
changes are needed at block boundaries. A deletion from the core
of an element that is also in the buffer, a removal from the buffer
of an element that is also in the core, or an insertion of an element
already held, does not change $S_t$ and costs no recourse.

The held sets satisfy $S_t\subseteq X_t$ and $|S_t|\le k$. Every core
set visited by the stored sequence
has at most $\kappa+D$ elements by~\eqref{eq:transition-value}, and the
buffer $H_t\subseteq X_t\setminus X_{t_{j-1}}$ has at most $2b$ elements, so
$|S_t|\le k$ by~\eqref{eq:det-capacity}. All elements of the core are in
$X_{\tau_i}\subseteq X_t$, so $S_t\subseteq X_t$.

We now bound the value, first outside the core move. Let $t\in B_{j+1}$, $j\ge1$.
Before the move, $S_t\supseteq A_{j-1,\kappa}\cup R$ with
$R=X_t\setminus X_{t_{j-1}}$, a prefix of length $t-t_{j-1}\le2b$ supplied
at $t_{j-1}$. Theorem~\ref{thm:prefix} and
Lemma~\ref{lem:prefix-benchmark} give $f(S_t)\ge a_\kappa\OPT^\kappa_t$
for $j\ge2$, and $S_t\supseteq X_t$ gives $f(S_t)\ge\OPT^\kappa_t$ for $j=1$.
After the core reaches $A_{j,\kappa}$, including during the buffer
removals, $S_t\supseteq A_{j,\kappa}\cup(X_t\setminus X_{t_j})$ and the
same argument at $t_j$ gives $f(S_t)\ge a_\kappa\OPT^\kappa_t$. By
Lemma~\ref{lem:curve} and $1/\kappa\le\eta$, $a_\kappa\ge q-\eta$.

During the core move, suppose first that the chosen
$i$ satisfies~\eqref{eq:quiet-interval}, and let
$t\in(\tau_i,\tau_{i+1}]$. The buffer contains
$Q_{\tau_i}=X_{\tau_i}\setminus X_{t_{j-1}}$, because no buffer element is
deleted before the core move ends. So, for the current core set $\Gamma$,
Lemma~\ref{lem:time-bounds} gives
\[
 f(S_t)\ge F_{\tau_i}(\Gamma)\ge(q-5\eta)E_{\tau_i}\ge(q-5\eta)V_i.
\]
By Lemma~\ref{lem:time-bounds} and~\eqref{eq:quiet-interval},
$\OPT^\kappa_t\le\OPT^\kappa_{\tau_{i+1}}\le V_{i+1}\le(1+\eta)V_i$. Hence
\[
 f(S_t)\ge\frac{q-5\eta}{1+\eta}\OPT^\kappa_t\ge(q-6\eta)\OPT^\kappa_t,
\]
where the last inequality is equivalent to
$q-5\eta\ge q-6\eta+q\eta-6\eta^2$, which holds because $q<1$.

If no interval is quiet, then
$V_{v+1}>(1+\eta)V_v$ for every $v<N$, and $i=N-1$. Applying this
inequality for $v=0,\ldots,N-2$ gives $V_{N-1}\ge(1+\eta)^{N-1}V_0$.
Since $N-1\ge8/\eta$, Bernoulli's inequality
$(1+\eta)^{N-1}\ge1+(N-1)\eta$ gives $V_{N-1}\ge9V_0$. By Lemma~\ref{lem:time-bounds},
\[
 \OPT^\kappa_{\tau_{N-1}}\ge V_{N-1}/2\ge4.5V_0\ge4\OPT^\kappa_{t_j}.
\]
Let $t\in(\tau_{N-1},\tau_N]$ and $H_t=X_t\setminus X_{t_j}$, which is in
the buffer. If $O$ attains $\OPT^\kappa_t$, then $O\cap X_{t_j}$ has at
most $\kappa$ elements, so~\eqref{eq:subadditive} and monotonicity give
$\OPT^\kappa_t\le f(O\cap X_{t_j})+f(O\cap H_t)\le\OPT^\kappa_{t_j}+f(H_t)$.
Since $\OPT^\kappa_t\ge\OPT^\kappa_{\tau_{N-1}}\ge4\OPT^\kappa_{t_j}$,
\[
 f(S_t)\ge f(H_t)\ge\OPT^\kappa_t-\OPT^\kappa_{t_j}
 \ge\tfrac34\OPT^\kappa_t\ge(q-6\eta)\OPT^\kappa_t.
\]

The last three paragraphs give
$f(S_t)\ge(q-6\eta)\OPT^\kappa_t$ at every time. By
Lemma~\ref{lem:cardinality}, $\OPT^\kappa_t\ge(\kappa/k)\OPT_t$, and
$\kappa/k\ge1-3\eta-1/k\ge1-4\eta$. Therefore
\[
 f(S_t)\ge(q-6\eta)(1-4\eta)\OPT_t\ge(q-10\eta)\OPT_t
 =(q-\eps_0/2)\OPT_t\ge(q-\eps)\OPT_t.
\]

For the recourse, note that the chosen interval contains at least $w$
updates, and the operation list has at most $4\kappa+b\le5k$ entries
(Lemma~\ref{lem:transition} and $|B_j|=b$). Each update performs at most
$\lceil5k/w\rceil$ operations from the list and inserts its arrival, so
by~\eqref{eq:det-interval-length},
\begin{equation}
 C_t\le\lceil180/\eta^2\rceil+1=O(\eta^{-2})\qquad\text{when }k\ge K_\eta.
 \label{eq:det-large-recourse}
\end{equation}
Outside the interval, each update only inserts its arrival. Together with
the bound $2k<2K_\eta$ for $k<K_\eta$,
$C_t=O(\eta^{-3})=O(\eps^{-3})$, since
$\eta\ge\eps/40$.

As for the lookahead, all computations for block
$B_{j+1}$ happen at $t_j$. Algorithm~1 at $t_j$ uses the arrivals up to
$t_j+2b$, and the candidate moves and $K_{\tau_N}$ use arrivals up to
$t_{j+1}=t_j+b$. So the lookahead is $2b\le2\eta k=\eps_0k/10\le
\lceil\eps k/3\rceil$. Algorithm~1 queries sets of at most $\kappa+2b$
elements by Theorem~\ref{thm:prefix}. Each query in
Lemma~\ref{lem:transition} is $F_\tau(S)=f(Q_\tau\cup S)$ with
$|S|\le\kappa$ and $|Q_\tau|\le2b$, and the ratios
in~\eqref{eq:initial-transition-bound} query sets of the same form. The
greedy runs for $K_\tau$ query sets of at most $\kappa$ elements. By
\eqref{eq:det-capacity}, every query has at most $k$ elements.

Finally, there are at most
$n/b+3$ checkpoints. At each, Algorithm~1 uses polynomially many queries,
and there are $N=O(1/\eta)$ candidate starts, each with $O(1/\eta)$
attempts by Lemma~\ref{lem:time-bounds}. Each attempt makes
$O(m\kappa(|X_\tau|+\kappa+1)^D)\le(n+1)^{O(D)}$ queries by
Lemma~\ref{lem:transition}, where we count dummy elements in the ground
set and use $\kappa\le k<n$, $m=O(1/\eta)$ and
$D=O(\eta^{-2})$. Since $n\ge k\ge4D/\eta$, the factors
polynomial in $n$, $k$ and $1/\eta$ are at most $(n+1)^{O(1)}$, and the
total is $(n+1)^{O(\eta^{-2})}=(n+1)^{O(\eps^{-2})}$. The other computation consists of the tests
in~\eqref{eq:residual-test}, the minima~\eqref{eq:initial-transition-bound}
and~\eqref{eq:time-suffix}, and bookkeeping, and is bounded in the same
way.
\end{proof}

\subsection{Proof of Corollary~\ref{cor:full-foresight}}
\label{app:full-foresight}

We use the following consequence of \citet[Theorems~5 and~6]{balkanski2026}:
for every instance with foresight, there is a sequence of sets
$S_t\subseteq X_t$ with $|S_t|\le k$, $f(S_t)\ge\frac23\OPT_t$ for every
$t$, and $|S_t\triangle S_{t-1}|\le C_0$ for every $t$, where $C_0$ is an
absolute constant. \Citet[Theorem~6]{balkanski2026} with accuracy $1/100$
gives ratio
$1/\sqrt2-1/100>2/3$.

\begin{proof}
If $k\ge K_\eta$, use the algorithm of
Theorem~\ref{thm:deterministic}, whose recourse is $O(\eta^{-2})$
by~\eqref{eq:det-large-recourse}. Assume
$k<K_\eta=O(\eta^{-3})$. Query
every subset of $\{e_1,\ldots,e_n\}$ with at most $k$ elements; there are
at most $(n+1)^k$. From these values compute $\OPT_t$ and the sets
\[
 \mathcal V_t=\{S\subseteq X_t:|S|\le k,\ f(S)\ge\tfrac23\OPT_t\}
 \qquad(1\le t\le n),
\]
and put $\mathcal V_0=\{\varnothing\}$.

For $S\in\mathcal V_t$, let $c_t(S)$
be the smallest possible value of $\max_{1\le u\le t}|S_u\triangle
S_{u-1}|$ over sequences $S_0=\varnothing$, $S_u\in\mathcal V_u$ for
$1\le u\le t$, with $S_t=S$. Then $c_0(\varnothing)=0$ and
\[
 c_t(S)=\min_{S'\in\mathcal V_{t-1}}\max\{c_{t-1}(S'),|S\triangle S'|\}.
\]
Indeed, every such sequence ending at $S$ has a last but one set
$S'\in\mathcal V_{t-1}$, and its largest change is the larger of the
largest change of the sequence up to $S'$ and $|S\triangle S'|$;
minimizing first over sequences ending at $S'$ and then over $S'$ gives
the recurrence. The algorithm computes all $c_t$, stores a minimizing
$S'$ for each $S$ (ties broken by a fixed order), picks a minimizer of
$c_n$ on $\mathcal V_n$, and follows the stored predecessors back to
time $0$.

The returned sequence has $S_t\in\mathcal V_t$,
so $S_t\subseteq X_t$, $|S_t|\le k$ and
$f(S_t)\ge\frac23\OPT_t\ge(q-\eps)\OPT_t$. The
sequence from the cited result is one of the sequences in the minimum, so
the returned recourse is at most $C_0$. Every query has at most $k$
elements. The dynamic program examines at most $(n+1)^{2k}$ pairs per
layer and $n$ layers, so both the number of queries and the
number of other operations are $(n+1)^{O(k)}=(n+1)^{O(\eta^{-3})}$.
Combining the two cases gives recourse
$\max\{C_0,O(\eta^{-2})\}=O(\eps^{-2})$ and
$(n+1)^{O(\eps^{-3})}$ queries.
\end{proof}

\subsection{Proof of Theorem~\ref{thm:randomized}}
\label{app:randomized-theorem}

\begin{proof}
We analyze Algorithm~4, with $\theta=\eps_0/6\le1/12$. If
$k<8\theta^{-2}$, Lemma~\ref{lem:base-greedy} gives $f(S_t)\ge q\OPT_t$,
two sets of at most $k$ elements differ in at most $2k<16\theta^{-2}$
elements, and the algorithm uses no lookahead. Greedy on $X_t$ with
budget $k$ makes $O(kt)$ queries, so this case makes $O(kn^2)$ queries in
total. Suppose $k\ge8\theta^{-2}$. Then
$\kappa=k-2\lfloor\theta k\rfloor\ge(1-2\theta)k\ge\tfrac34k\ge
6\theta^{-2}$, so $1/\kappa\le\theta$. By Theorem~\ref{thm:prefix} and Lemma~\ref{lem:curve}, the
output of Algorithm~1 is $(q-1/\kappa)$-addition-robust, and
$q-1/\kappa\ge q-\theta$. Lemma~\ref{lem:checkpoint} with $\alpha=q-\theta$ gives
\[
 \E f(S_t)\ge(q-4\theta)\OPT_t\ge(q-\eps)\OPT_t
\]
at every fixed time, and recourse at most $2\lceil8/\theta^2\rceil+1$
for every outcome. Since $\eps_0\ge\eps/2$, both cases have recourse
$O(\eps^{-2})$.

For the probability bound, fix $t$. If $k<8\theta^{-2}$, then
$f(S_t)\ge q\OPT_t$ always. Otherwise, during $B_1$ the held set is
$X_t$. For $t$ in a block $B_{j+1}$ with $j\ge1$, the window chosen for
this block contains $t$ with probability $1/n_w\le\theta\le\eps/6$. If
$t$ lies before the chosen window, then
$S_t\supseteq A_{j-1,\kappa}\cup(X_t\setminus X_{t_{j-1}})$, and if it
lies after it, then $S_t\supseteq A_{j,\kappa}\cup(X_t\setminus X_{t_j})$;
for $j=1$ the first case gives $S_t=X_t$. In both cases the arrivals
since the checkpoint form a prefix that was supplied to Algorithm~1 at
that checkpoint, so Theorem~\ref{thm:prefix},
Lemma~\ref{lem:prefix-benchmark} and Lemma~\ref{lem:cardinality} give
\[
 f(S_t)\ge\Bigl(q-\frac1\kappa\Bigr)\frac\kappa k\OPT_t
 \ge(q-\theta)(1-2\theta)\OPT_t\ge(q-3\theta)\OPT_t,
\]
where the last step uses $(q-\theta)(1-2\theta)=q-(2q+1)\theta+2\theta^2$
and $2q+1<3$. Since $3\theta=\eps_0/2\le\eps$, this gives
$\Pr\{f(S_t)\ge(q-\eps)\OPT_t\}\ge1-\eps/6$.

There are at most $n/b+3$ checkpoints. Each calls Algorithm~1 with
budget $\kappa\le k$, at most $n+k$ real or dummy old elements and
$J=2b+1\le3b$ prefixes, so by Theorem~\ref{thm:prefix} it makes
$O(\kappa(n+2k)J)=O(knb)$ queries, using $k<n$. The schedule makes no
further queries, so the total is $O((n/b+3)knb)=O(kn^2)$, since
$b\le k<n$. Every query has at most $\kappa+2\lfloor\theta k\rfloor
=k$ elements, and the only future elements queried are among the next
$2\lfloor\theta k\rfloor\le\eps k/3$ arrivals. At the end of the sequence,
zero-value padding arrivals are appended as in the proof of
Theorem~\ref{thm:deterministic}.
\end{proof}

\subsection{Proof of Proposition~\ref{prop:oracle-upper}}
\label{app:oracle-upper}

We use the static bound in the following form
\citep{nemhauserwolsey1978}, \citep[Theorem~1.8 and Section~2]{vondrak2013}:
for every constant $\alpha_0>q$, no randomized algorithm that makes
polynomially many value queries returns, for every nonnegative monotone
submodular $g$ on a ground set $V$ and every $k$, a set $S\subseteq V$ with
$|S|\le k$ and $g(S)\ge\alpha_0\max_{|O|\le k}g(O)$ with probability at
least $2/3$. The queries may be arbitrary subsets of $V$.

\begin{proof}
A guarantee with some
$\upsilon$ implies the guarantee with every smaller $\upsilon$, so we may
assume $0<\upsilon<1/e$, and put $\alpha=q+\upsilon<1$. Suppose, for
contradiction, that an algorithm $\mathcal A$ makes polynomially many
queries and satisfies $\E f(S_t)\ge\alpha\OPT_t$ at every fixed time for
every normalized monotone submodular $f$ and every arrival order.

We use $\mathcal A$ to build a static algorithm. Let $g$ be nonnegative, monotone and submodular
on $V=\{v_1,\ldots,v_n\}$, and let $k$ be given. Query
$c=g(\varnothing)$ and put $f(S)=g(S)-c$. Then $f(\varnothing)=0$, $f\ge0$
by monotonicity of $g$, and $f$ is monotone and submodular. Run
$\mathcal A$ on the arrival order $v_1,\ldots,v_n$ with budget $k$,
answering each query $f(S)$ by the query $g(S)$ minus $c$; this works for
every $S\subseteq V$, including sets with future elements. The final
output satisfies $S_n\subseteq V$ and $|S_n|\le k$.

Let $\OPT_g=\max_{|O|\le k}g(O)$, so
$\OPT_n=\OPT_g-c$. The guarantee at time $n$ and $c\ge0$ give
\[
 \E g(S_n)=\E f(S_n)+c\ge\alpha(\OPT_g-c)+c
 =\alpha\OPT_g+(1-\alpha)c\ge\alpha\OPT_g.
\]

If $\OPT_g=0$, any output
is optimal. Otherwise let $Y=g(S_n)/\OPT_g\in[0,1]$ and
$\alpha_0=q+\upsilon/2$. Since $Y<\alpha_0$ on the event $\{Y<\alpha_0\}$
and $Y\le1$ always,
$\alpha\le\E Y\le\alpha_0+(1-\alpha_0)\Pr\{Y\ge\alpha_0\}$, so
\[
 \Pr\{Y\ge\alpha_0\}\ge\frac{\alpha-\alpha_0}{1-\alpha_0}
 =\frac{\upsilon/2}{1-\alpha_0}\ge\frac\upsilon2.
\]
Run the simulation independently $N_0=\lceil2\ln3/\upsilon\rceil$ times
and return the output with the largest value of $g$. All runs stay below
$\alpha_0\OPT_g$ with probability at most
$(1-\upsilon/2)^{N_0}\le e^{-\upsilon N_0/2}\le1/3$. Since $\upsilon$ is
fixed, the total number of queries is polynomial, and this static
algorithm contradicts the bound stated before the proof, because
$\alpha_0>q$.

The simulation places no restriction
on consecutive outputs, so the contradiction holds with unrestricted
recourse. The order $v_1,\ldots,v_n$ and the time $n$ are fixed before
$\mathcal A$ draws its random bits, as the fixed-time guarantee requires.
\end{proof}

\subsection{Proof of Proposition~\ref{prop:lookahead-lower}}
\label{app:lookahead-lower}

We use the barrier of \citet{fu2026} for algorithms without lookahead,
which, as in \citet[Section~2.1]{fu2026}, do not know the length of the
stream in advance.

\begin{lemma}[{\citealp[Theorem~1.1(ii) and Theorem~5.1]{fu2026}}]
\label{lem:fu-barrier}
For every $\zeta>0$ there are constants $c_\zeta,d_\zeta>0$ such that the
following holds for every sufficiently large integer $k$. Let a randomized
algorithm with budget $k$ query only sets of arrived elements, and suppose
that on every instance with at most $d_\zeta k$ elements and for every
outcome of its random choices it makes at most $e^{c_\zeta k}$ value
queries before the last arrival and satisfies
$|S_t\triangle S_{t-1}|\le c_\zeta k$ for every $t$. Then some instance
with at most $d_\zeta k$ elements, fixed before the random choices, has
$\E f(S_t)<(2-\sqrt2+\zeta)\OPT_t$ at its last arrival $t$.
\end{lemma}

\begin{proofof}{Proposition~\ref{prop:lookahead-lower}}
Put $\zeta=\sigma-(2-\sqrt2)>0$ and take $c_\zeta,d_\zeta$ from
Lemma~\ref{lem:fu-barrier}. Fix $k$, write $\Lambda=\Lambda_k$, and
let $\mathcal A$ be the algorithm for budget $k$. If $C_k=0$, the output is
always empty, which violates the guarantee on any instance with
$\OPT_1>0$; if $\Lambda>k$, then $C_k(\Lambda+1)>k$. So assume $C_k\ge1$ and $\Lambda\le k$.

Pad the arrival sequence by placing $\Lambda$ new elements directly after each
arrival $e_i$, extend $f$ to them by ignoring them, so that they have
value zero, and let $\pi(i)$ be the position of $e_i$ in the padded
sequence. Since $\mathcal A$ knows only the next $\Lambda$ padded arrivals
(Section~\ref{sec:model}), at time $\pi(i)$ it sees only padding, whether
or not $e_{i+1}$ exists, and at the $\Lambda$ padded times after $\pi(i)$ it
sees nothing beyond $e_{i+1}$ and padding. An algorithm $\mathcal B$
without lookahead can therefore, when $e_i$ arrives, run $\mathcal A$ on
the padded steps $\pi(i-1)+1,\ldots,\pi(i)$, answer each query by querying
its real elements, all of which are among $e_1,\ldots,e_i$, and output
$S^{\mathcal B}_i=S^{\mathcal A}_{\pi(i)}\cap\{e_1,\ldots,e_i\}$.

The optimum of the padded instance at time $\pi(i)$ is $\OPT_i$, so
$\E f(S^{\mathcal B}_i)\ge\sigma\OPT_i$ at every fixed time. Since
$\mathcal A$ makes $\Lambda+1$ updates between two outputs of $\mathcal B$,
$|S^{\mathcal B}_i\triangle S^{\mathcal B}_{i-1}|\le C_k(\Lambda+1)$. On an
instance with at most $d_\zeta k$ elements the padded instance has at
most $d_\zeta k(k+1)$ elements, so $\mathcal B$ makes at most
$p(d_\zeta k(k+1))\le e^{c_\zeta k}$ queries when $k$ is large. If also
$C_k(\Lambda+1)\le c_\zeta k$, Lemma~\ref{lem:fu-barrier} gives an instance with
$\E f(S^{\mathcal B}_t)<\sigma\OPT_t$ at its last arrival $t$, a
contradiction. Hence $C_k(\Lambda_k+1)>c_\zeta k$ for all sufficiently
large $k$.
\end{proofof}

\bibliography{references}
\end{document}